\pdfoutput=1
\documentclass[twocolumn,UTF8,aps,prb,]{revtex4-1}
\usepackage{amsmath,amssymb,bm}
\usepackage{graphicx}
\usepackage{subfigure}
\usepackage{epstopdf}
\usepackage{xcolor}

\newcommand{\be}{\begin{equation}}
\newcommand{\ee}{\end{equation}}
\newcommand{\bea}{\begin{eqnarray}}
\newcommand{\eea}{\end{eqnarray}}
\newcommand{\bsube}{\begin{subequations}}
\newcommand{\esube}{\end{subequations}}

\begin{document}
\title{Dynamic Transport Characteristics of Side-Coupled Double Quantum-Impurity Systems}
\author{YiJie Wang$^{1}$}
\author{JianHua Wei$^{1}$} \email{wjh@ruc.edu.cn}
\affiliation {$^{1}$Department of Physics, Renmin University of China, Beijing 100872, China}
\begin{abstract}
A systematic study is made on the time-dependent dynamic transport characteristics of the side-coupled double quantum-impurity system based on the hierarchical equations of motion. It is found that the transport current behaves like a single quantum dot when the coupling strength is low during tunneling or coulomb coupling. The dynamic current oscillates due to the temporal coherence of the electron tunneling device only when the tunneling transition is coupled. The oscillation frequency of the transport current is related to the step voltage applied by the lead, while the $T$, e-e interaction $U$ and the bandwidth $W$ have little influence. The amplitude of the current oscillation exists in positive correlation with $W$ and negative correlation with $U$. With the increase in coupling $t_{12}$ between impurities, the ground state of the system changes from a Kondo singlet of one impurity to a spin-singlet of two impurities. Moreover, lowering the temperature could promote the Kondo effect to intensify the oscillation of the dynamic current. When only the coulomb transition is coupled, it is found that the two split-off Hubbard peaks move upward and have different interference effects on the Kondo peak at the Fermi surface with the increase in $U_{12}$, from the dynamics point of view. 
\end{abstract}
\maketitle
\section{\bf{Introduction}}
Quantum dots can form different quantum-impurity systems with different leads as a typical low-dimensional mesoscopic system (metal lead, ferromagnetic lead, superconducting lead, etc.), which serves as a more detailed research tool in the field of strong correlation. Since the double quantum-impurity system possesses a variety of geometrical configurations, its tunneling path is more than that of the single quantum-impurity system, and thus there are more abundant physical properties belonging to the double quantum impurity systems. The Aharonov-Bohm oscillation, Fano resonance, Kondo effect, quantum phase transition, thermoelectric effect are a few examples\cite{1,2,3,4,5,6,7,8,9,10}. The research on the transport properties of the double quantum-impurity systems not only provides a theoretical basis for the study of integrated circuits but also plays a vital role in the study of quantum bit, quantum regulation, and other aspects \cite{11}.\\
\indent
Although many studies have been made on the spatial coherence of the electron wave function in quantum-impurity systems, the temporal coherence of the same remains to be studied due to the difficulties involved in dealing with the time phase coherence and time memory effect of the electrons\cite{12,13}. There have been many studies on the transport properties of quantum-impurity systems. For example, for a system having no interaction, the traditional Landauer-Buttiker scattering matrix theory provides steady state current through the leads \cite{14,15}. Ned S. Wingreen et al. studied the time-dependent transport current by using the method of motion equation for the first time and gave the analytical expression of current by using Keldysh Green's function and Dyson's equation. Such oscillation behavior of time-dependent current has gathered people's attention \cite{16,17}. However, this method depends on the wide-band limit (WBL), which assumes that the energy band of the lead has no energy dependence and cannot be calculated quantitatively. Yu Zhu et al. used the time-domain decomposition method to directly calculate the time-dependent transport current I(t) in the numerical form by using Green's function. However, this method had difficulty in accurately solving the case of weak coupling between the device and the lead \cite{18}. Joseph Maciejko et al. theoretically provided an exact analytical expression for the transport current under nonequilibrium and nonlinear response conditions, but this expression needs to be derived once again after the step voltage applied to the leads is changed \cite{19}. From most studies made on the dynamic transport current of quantum-impurity systems, it can be seen that the electron-electron (e-e) interaction is either ignored or treated in the average field.\\
\indent
Some studies have been conducted on quantum-impurity systems despite having difficulty in the calculation of the time dependent transport current. For example, the density matrix renormalization group method is extended to be time-varying to deal with time-dependent one dimensional systems and the transport problems of single impurity systems \cite{20,21,22}. The time-dependent numerical renormalization group method was used for studying the nonequilibrium dynamics of the quantum-impurity system, and it was found that the occupancy number would appear as Rabi oscillation when the time-dependent gate voltage was applied on the quantum-impurity \cite{23,24}. However, the use of perturbation processing has not been able to accurately describe the latest physical phenomena, and the above work focuses on single-level resonant tunneling, which does not aptly explain the interesting phenomenon of time-dependent transport characteristics in the Kondo system.\\
\indent
In this paper, a general approach is proposed based on the hierarchical equations of motion (HEOM) form to investigate the nonequilibrium dynamics of quantum impurity systems while taking into account the environmental effects. The time-dependent quantum transport problem is solved by using a series of equations of motion to calculate the time-dependent transport current of the quantum-impurity systems with the help of a non-perturbed quantum impurity model. These are all related to the experiments on quantum dots and quantum wires, which are of great significance for quantum computation in the above-mentioned nanometer devices.\\
\indent
Presently, our group has done some research based on this method. One is related to the tunneling coupling between quantum dots. Yongxi Cheng pointed out that the temperature inhibits the oscillation of dynamic current by inhibiting kondo effect due to the temporal coherence of electrons for the single quantum dot system \cite{25}. For the parallel-coupled double quantum dots system, the current oscillation is similar to that of the single quantum dot, and different coupling strengths have different forms of oscillation. This research is based on the perspective of dynamics.
\\
\indent
The other aspect is regarding the coulomb transition coupling between the quantum dots. Fuli Sun pointed out that the different coulomb coupling strength between the side-coupled double quantum dots divided the singly-occupied (S-O) state of quantum dot 1 into three quasi-particle substates from the perspective of the spectral function \cite{4}. The spectral functions show different characteristics in different kondo regions. The effects of different coupling on different models were separately studied. Based on their work, this paper further studies the effects of the two coupling modes on the side-coupling double quantum dots system.\\
\indent
The structure of this article is organized as follows. In part 2, the HEOM method is introduced and derived, and the common form of the time-dependent quantum transport current in a quantum-impurity system is given. In part 3, the influences of the different step voltage V, tunneling transition coupling $t_{12}$ temperature T, e-e interaction U, and bandwidth W on the transport current within the side-coupled double quantum dots system are studied. Secondly, we also study the influence of the Coulomb interaction coupling U on the transport current of the side-coupled double quantum dots system in different Kondo regions. Part 4 provides the summary of the work.
\section{\bf{Model and Hamiltonian}}
\indent
Based on the HEOM equation, we developed a set of non-perturbation methods for solving quantum impurities \cite{26,27,28,29}, which can not only solve the quantum impurities problem in open systems, but also deal with the quantum dissipation problem in non-equilibrium.\\
\begin{figure}[htbp]
\centering
\includegraphics[width=0.4\textwidth]{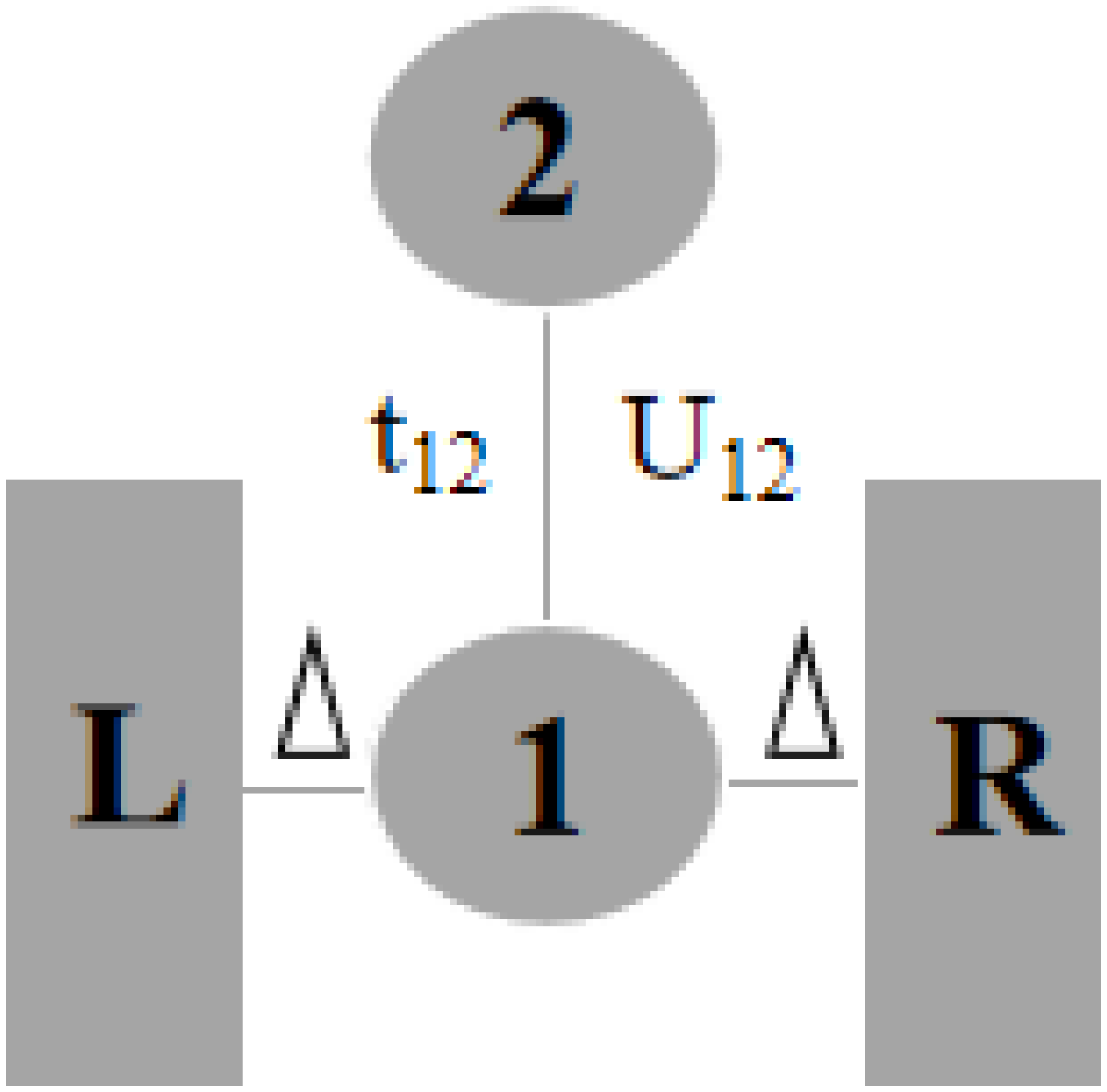}
\caption{Side-coupled double quantum-impurity systems model.}
\end{figure}
\indent
As shown in figure 1, we studied the side-coupled double quantum dots model with particle-hole symmetry ($\varepsilon\uparrow$=$\varepsilon\downarrow=-U/2$). QD1 and QD2 are identical, in which the two sides of QD1 are connected to the left and right leads respectively, and there may be tunneling transition coupling $t_{12}$ and coulomb interaction coupling $U_{12}$ between QD2 and QD1. We take two quantum dots as research objects and describe their physical properties through the motion equation of the density matrix operator. The leads are regarded as the environment attached to the quantum dots, which is generally considered as a non-interacting fermi library. The influence of the environment on the quantum dots is described by the correlation function. By constructing a set of non-perturbative equations of motion about the reduced density matrix operator and the auxiliary density matrix operator, the dynamic transport current of the system is solved in the HEOM linear space. Under the Anderson impurity model, the total Hamiltonian of the quantum dots system, the leads and the interaction between them is\begin{equation}
H=H_{S}+H_{B}+H_{int}.
\end{equation}
\indent
The Hamiltonian of the quantum dot system is:
\begin{equation}
H_{S}=\sum_{i\sigma}\varepsilon_{i\sigma}\hat{a}^{+}_{i\sigma}\hat{a}_{i\sigma}+\frac{U}{2}\sum_{i\sigma}n_{i\sigma}n_{i\bar{\sigma}}+\gamma\sum_{<i,j>\sigma}(\hat{a}^{+}_{i\sigma}\hat{a}_{j\sigma}+\hat{a}^{+}_{j\sigma}\hat{a}_{i\sigma})
\end{equation}
wherein, $\varepsilon_{i\sigma}$ refers to the on-site energy of the electron with spin $\sigma(\sigma=\uparrow,\downarrow)$ in the quantum dot i(i=1,2), $\hat{a}^{+}_{i\sigma}$ and $\hat{a}_{i\sigma}$ correspond to the creation and annihilation operators of electrons with spin $\sigma$ on quantum dot i, $n_{i\sigma}=\hat{a}^{+}_{i\sigma}\hat{a}_{i\sigma}$ is the electron number operator of quantum dot i, $U$ is the coulomb interaction between the electrons of spin $\sigma$ and $\bar{\sigma}$ in the quantum dot, $\gamma$ refers to the coupling strength between quantum dots i and j.\\
\indent
The Hamiltonian of the leads is
\begin{equation}
H_{B}=\sum_{k,\mu,\alpha=L,R}\varepsilon_{k\alpha}\hat{d}^{+}_{k\mu\alpha}\hat{d}_{k\mu\alpha},
\end{equation}
where $\varepsilon_{k\alpha}$ refers to the energy of electrons with the wave vector k at $\alpha$ leads, $\mu$ is the orbital of the electron(including spin orbital $\sigma$ and space orbital i), $\hat{d}^{+}_{k\mu\alpha}$ and $\hat{d}_{k\mu\alpha}$ are the corresponding creation and annihilation operators.\\
\indent
The Hamiltonian coupling of quantum dots system and leads is 
\begin{equation}
H_{int}=\sum_{\mu}[f^{+}_{\mu}(t)\hat{a}_{\mu}+\hat{a}^{+}_{\mu}f_{\mu}(t)],
\end{equation}
where $f^{+}_{\mu}(t)=e^{iH_{B}t}[\sum_{k\alpha}t^{*}_{\alpha k\mu}\hat{d}^{+}_{\alpha k\mu}]e^{-iH_{B}t}$ is the random interaction operator satisfying Gaussian statistics, $t_{\alpha k\mu}$ is the transfer coupling matrix element. We describe the effect of fermik library on the quantum dots system through the hybrid function, and here we consider using Lorentz form, $\Delta_{\alpha}(\omega)=\pi\sum_{k}t_{\alpha k\mu}t^{*}_{\alpha k\mu}\delta(\omega-\varepsilon_{k\alpha})=\Delta W^{2}/[2(\omega-\mu_{\alpha})^{2}+W^{2}]$ ,where $\Delta$ is the effective coupling strength between quantum dots and leads, $W$ is the bandwidth of the leads, $\mu_{\alpha}$ is the chemical potential of lead $\alpha$ \cite{29,30}.\\
\indent
The closed HEOM equation of the open system is
\begin{align}
\dot\rho^{n}_{j_1\cdots j_n} =& -\Big(i{\cal L} + \sum_{r=1\cdots n}\gamma_{j_r}\Big)\rho^{n}_{j_1\cdots j_n}
-i \sum_{j}\!
{\cal A}_{\bar j}\, \rho^{n+1}_{j_1\cdots j_nj}\nonumber \\
& -i \sum_{r=1\cdots n}(-)^{n-r}\, {\cal C}_{j_r}\,
\rho^{n-1}_{j_1\cdots j_{r-1}j_{r+1}\cdots j_n},
\end{align}
wherein,
\begin{align}
\!
{\cal A}_{\bar j}\rho^{n+1}_{j_1\cdots j_n j} =a^{\bar{o}}_{\mu}\rho^{n+1}_{j_1\cdots j_n j}+(-)^{n+1}\rho^{n+1}_{j_1\cdots j_n j}a^{\bar{o}}_{\mu},
\end{align}
\begin{align}
{\cal C}_{j_r}\rho^{n-1}_{j_{1r}\cdots j_{nr}} =\sum_{\nu}[ {\cal C}^{o}_{\alpha\mu\nu}a^{o}_{\nu}\rho^{n-1}_{j_{1r}\cdots j_{nr}}-(-)^{n-1}{\cal C}^{\bar{o}}_{\alpha\mu\nu}\rho^{n-1}_{j_{1r}\cdots j_{nr}}a^{o}_{\nu}].
\end{align}
The index $j=\{\alpha,\mu,\sigma\}$ is associated with characteristic memory time $\gamma^{-1}_{m}$. $\{\rho^{n}_{j_1\cdots j_n} (t),(n=1\cdots$L)\} denotes the auxiliary density operator of order n, L denotes the truncation order. $\cal L,\cal A, \cal C$ represent superoperators, as defined in [27]. $a^{o}_{\nu}$ and $a^{\bar{o}}_{\nu}$ represent the creation and annihilation operators of electrons with the electron orbital of $\mu(o=+/-)$. The correlation function ${\cal C}^{o}_{\alpha\mu\nu}(t-\tau)=<f^{o}_{\alpha\mu}(t)f^{\bar{o}}_{\alpha\mu}(\tau)>_{B}$ follows the time-reversal symmetry and detailed balance relations.\\
\indent
Let's start from the initial equilibrium steady state of the system:
\begin{align}
\mu_{\alpha}=\mu^{eq}=0.
\end{align} 
When the voltage is applied to the leads, the system is out of balance and the current flowing from the left to the right lead is $I(t)=I_{L}(t)=I_{R}(t)$. The electron occupancy number $N(t)$ of the quantum dots is expressed as $N(t)=tr[a^{+}_{i\mu}a_{i\mu}\rho(t)]$. The current flowing into the quantum dots through lead $\alpha$ is expressed as
\begin{align}
I_{\alpha}(t)=-e\frac{d}{dt}<N_{\alpha}>_{T}=e\frac{i}{\hbar}<[N_{\alpha},H(t)]>_{T}.
\end{align} 
$H(t)$ is the Hamiltonian of the whole complex system in the interaction representation. $N_{\alpha}=f^{+}_{\alpha i\mu}f_{\alpha i\mu}$ is the particle number operator of the lead. $<\cdots>_{T}$ is the statistical average of the entire complex system. Since $[N_{\alpha},H(t)]=[N_{\alpha},H_{int}(t)]$ and $tr_{T}=tr_{S}tr_{B}$ is used, the dynamic transport current flowing into leads is expressed as
\begin{align}
I_{\alpha}(t)=i\sum_{\mu}tr_{S}[\rho^{+}_{\alpha\mu}(t)\hat{a}_{\mu}-\hat{a}^{+}_{\mu}\rho^{-}_{\alpha\mu}(t)].
\end{align} 
$tr_{S}$ and $tr_{B}$ represent the trace of quantum dots system and lead respectively. $\rho^{+}_{\alpha\mu}=(\rho^{-}_{\alpha\mu})^{+}$ is the first-order auxiliary density operator obtained by solving HEOM equation. \\
\indent
As a numerical method, HEOM has the following characteristics. First of all, with the increase of truncation order L, the calculation results of the corresponding physical quantities gradually converge in the full energy domain. In our calculations, if the error of the numerical results for each element of the density matrix or spectral function matrix of L=N and L=N+1 is less than 5\%, then we consider the results to be convergent, since this will obtain sufficiently accurate values of the dynamic transport current. Secondly, HEOM has great advantages in solving quantum dots problems with high accuracy. In reference \cite{25}, the time-dependent dynamic current of single-level resonant tunneling obtained by the HEOM method is compared with the current obtained by the analytical formula based on Keldysh Green function, non-equilibrium Green function, time-dependent density matrix renormalization group and time-dependent numerical renormalization group, respectively. It is found that the HEOM method can not only describe the dynamic behavior of the single-level system well, but also is superior to the latest time-based numerical renormalization group method in numerical solution. This also provides the premise for us to study the transport properties in the Kondo region of the side-coupled double quantum dots system.
\section{\bf{Results and discussion}}
\indent
There may be two forms of interaction between the two quantum dots for the side-coupled double quantum-impurity system. One is the tunneling type, which is represented by the transition coupling parameter $t_{12}$ and is a result of the two-channel Kondo effect and the two-stage Kondo effect \cite{31,32,33,34}. 
The second is the capacitive type, which is described by the Coulomb repulsion seen between the dots. It is expressed by the coulomb interaction constant $U_{12}$. This interaction is involved in single electron switches and other transport phenomena \cite{35,36}. Since there are no energy levels present near the Fermi surface, the Kondo effect results in a large transport current and also the oscillating behavior of the current.\\
\indent
The Hamiltonian of the side-coupled double quantum dot system is
\begin{align}
H_{S}=\sum_{i\sigma=1,2}[\varepsilon_{i\sigma}\hat{a}^{+}_{i\sigma}\hat{a}_{i\sigma}+U_{i}n_{i\sigma}n_{i\bar{\sigma}}]+\gamma\sum_{\sigma}(\hat{a}^{+}_{1\sigma}\hat{a}_{2\sigma}+\hat{a}^{+}_{2\sigma}\hat{a}_{1\sigma}).
\end{align} 
The parameters are the same as mentioned above.\\
\indent
Firstly, the capacitive coupling is ignored, and only the tunneling transition coupling is dealt with. Since the Kondo resonance assists the tunneling of electrons at low temperatures to produce a large resonant transport current, a new type of current oscillation appears.\\
\begin{figure}[htbp]
\centering
\includegraphics[width=0.4\textwidth]{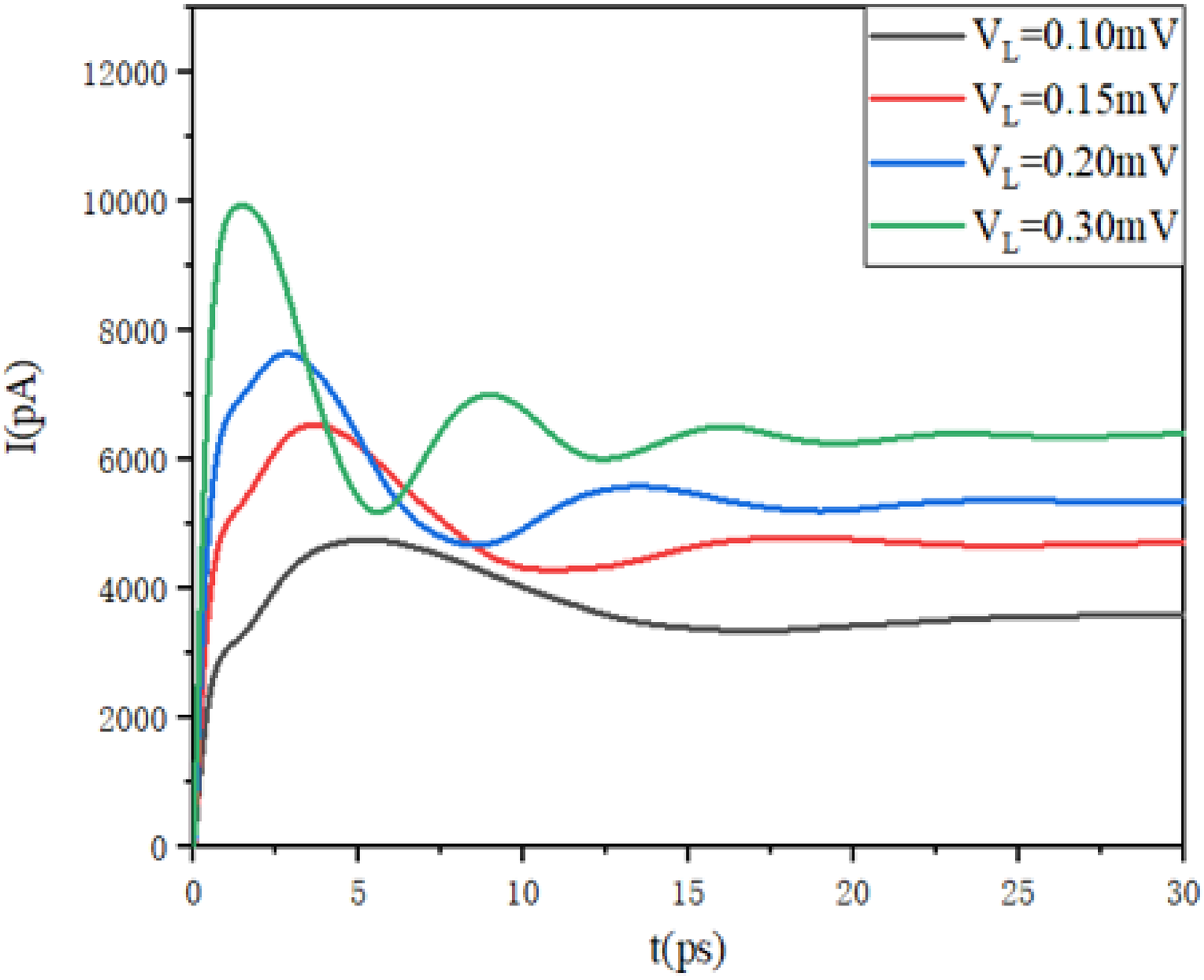}
\caption{The dynamic current of the side-coupled double quantum-impurity system at different step voltages.The parameters adopted are $t_{12}$=0.1meV,$K_{B}T$=0.015meV,$\Delta$=0.2meV,W=2meV,U=2meV,$U_{12}$=0, $\epsilon$$\uparrow$=$\epsilon$$\downarrow$=-1meV.}
\end{figure}
\indent
The current transport characteristics of the side-coupled double quantum-impurity system during different step voltages in the Kondo state are described in Figure 2. The tunneling coupling between the two impurities as weak coupling was set at 0. l meV. It could be seen that the current rapidly increases to its maximum value once the step voltage is applied, and further regular oscillations occur depending on the form of the step voltage applied. For example, when the voltage $V_{L}=-V_{R}=0.3mV$, the maximum value of current is 10000 pA, and three apparent oscillations occur before reaching its stable value. When $V_{L}=-V_{R}=0.1mV$, the maximum value of current is 5000 pA, and only one obvious oscillation occurs before reaching its stable value, and thus the amplitude is significantly reduced. This happens due to the time coherence corresponding to the step voltage as the electron tunnels through the quantum impurities. The accumulation and dissipation of the charges of the leads present on the left and right produce oscillatory behavior when the step voltage is changed suddenly. The larger the step voltage, the faster would be the accumulation and dissipation of the charge and a more obvious oscillation behavior. In addition, all of the current values reach their respective stable values at 30 ps, irrespective of the voltage pulse form, corresponding to their respective steady-state current.\\
\begin{figure}[htbp]
\centering
\includegraphics[width=0.4\textwidth]{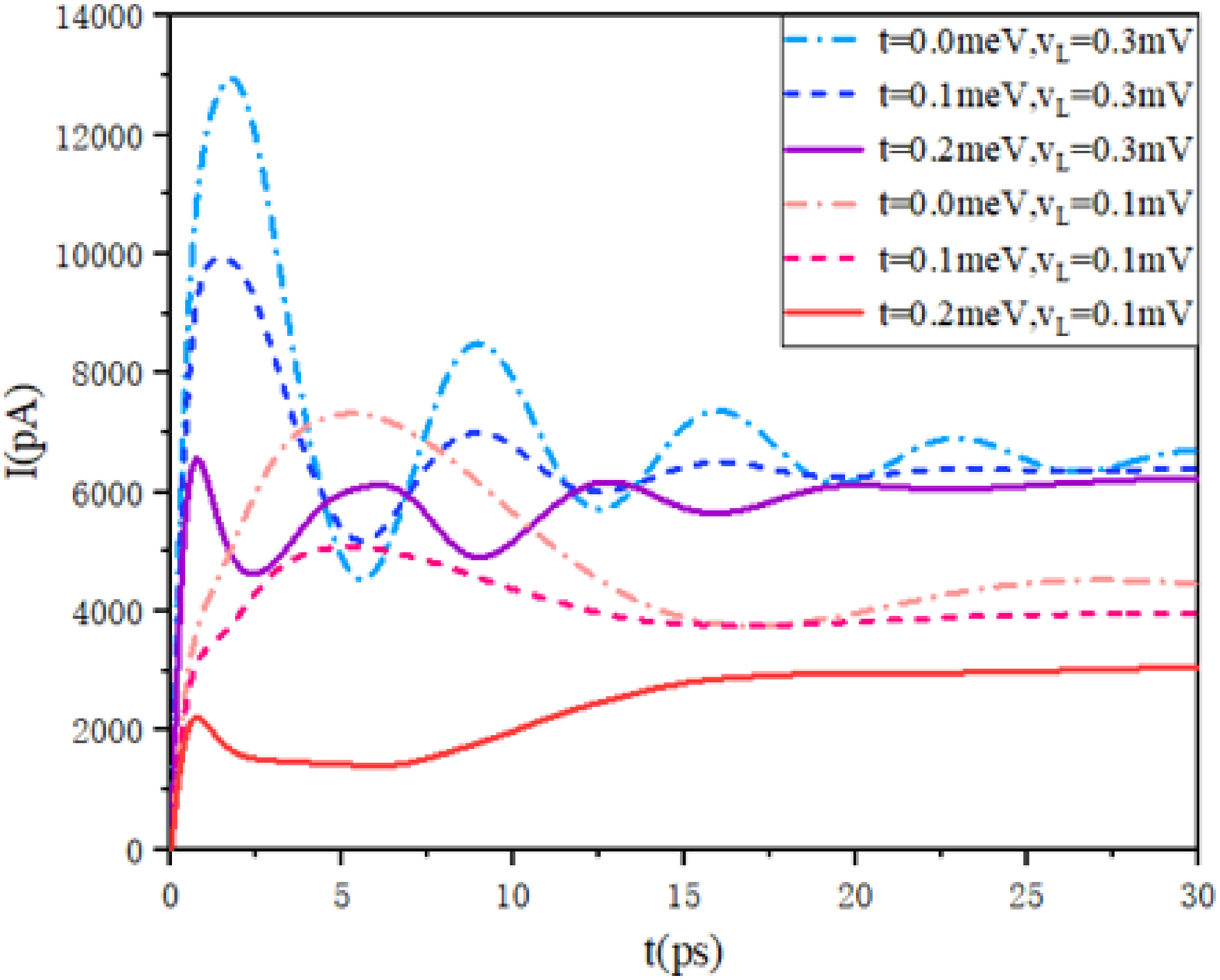}
\caption{The dynamic current of the side-coupled double quantum-impurity system at different step voltages and different transition coupling parameter $t_{12}$ .The parameters adopted are $K_{B}T$=0.015meV,$\Delta$=0.2meV,W=2meV,U=2meV,$U_{12}$=0, $\epsilon$$\uparrow$=$\epsilon$$\downarrow$=-1meV.}
\end{figure}
Figure 3 shows the oscillation form of the current in the side-coupled double quantum-impurity system at different $t_{12}$ and during large and small step voltages, respectively. It is found that, no matter how large a step voltage is added, the oscillation behavior of the current is similar to when $t_{12}=0$ and $t_{12}=0.1meV$, but the amplitude of the current oscillation, however, decreases and reaches a stable value faster when the coupling is weak. For example, when $V_{L}=-V_{R}=0.3mV$, the current of the system reaches a maximum value of 1.4 ps under the weak coupling condition of $t_{12}=0.1meV$ and reaches the steady-state current value of 20 ps. However, when $t_{12}=0$, the current reaches a maximum value of 1.8 ps, and there is still an oscillation behavior of small amplitude at 30 ps. This is because the ground state of the system maintains the Kondo singlet of each impurity when the tunneling transition coupling strength $t_{12}$ is weak, and thus exhibits an oscillating behavior similar to that of a single quantum dot. However, there will not only be an L-QD1-R for the current, but the L-QD1-QD2-QD1-R is also added to the system comparing it to the single quantum dot, which would speed up the current transport. As $t_{12}$ increases, the oscillation of the dynamic current changes significantly. When we compare the images of $t_{12}=0.1meV$ and $t_{12}=0.2meV$ under different step voltages, we find that no matter how big the step voltage is, when the system has weak coupling, the current would first reach a maximum value and then oscillate to a stable value, which is lower than the maximum value. However, when $t_{12}=0.2meV$ is in the strong coupling, the maximum value of the current that reaches first is the same as that of the steady-state value of the current that finally reaches after oscillation, or even slightly lower than the steady-state value itself. This is because when $t_{12}$ is small, the direct first-order coupling (t) is much stronger than the induced second-order antiferromagnetic spin coupling ($J= 4t^{2} /U$) present between the two impurities, in which case the current oscillates like a single quantum impurity. When $t_{12}$ is large, the spin-spin coupling $J$ dominates, making the ground state of the system transform into the spin-singlet of the two impurities, and thus the transport current shows different oscillatory behavior. \\
\begin{figure}[htbp]
\centering
\includegraphics[width=0.4\textwidth]{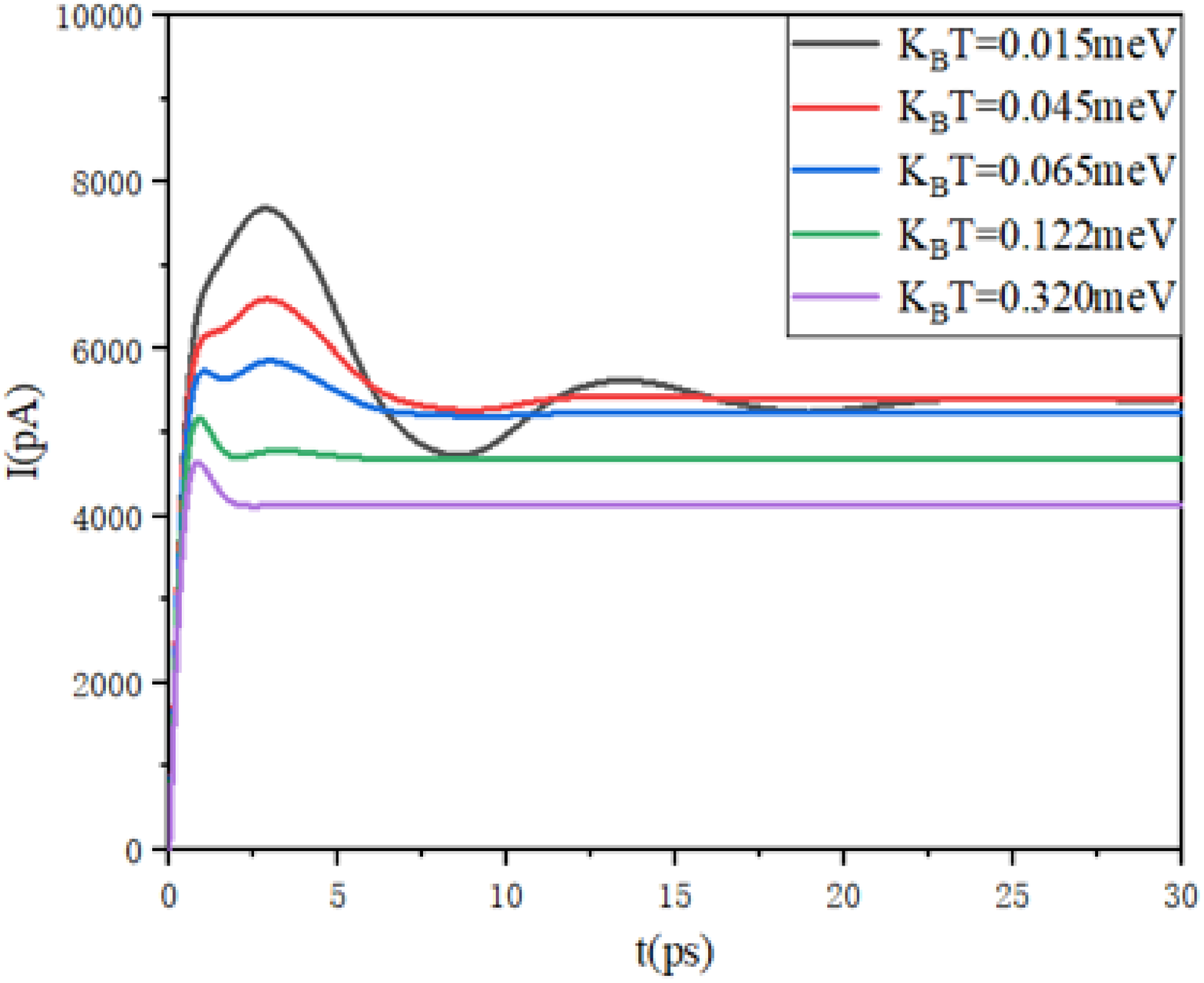}
\caption{The dynamic current of the side-coupled double quantum-impurity system at different temperature T.The parameters adopted are $t_{12}$=0.1meV,$V_{L}=-V_{R}=0.3mV$,$\Delta$=0.2meV,W=2meV,U=2meV,$U_{12}$=0,$\epsilon$$\uparrow$=$\epsilon$$\downarrow$=-1meV.}
\end{figure}
\indent In Figure 4, we have shown a graph of current changing with time at different temperatures. According to the formula
\begin{align}
T_{K}=\sqrt{\frac{U\tilde{\Delta}}{2}}e^{-\pi U/8\tilde{\Delta}+\pi\tilde{\Delta}/2U},
\end{align} 
where $\tilde{\Delta}=2\Delta$ ,the calculated Kondo temperature is approximately $K_{B}T\approx0.122meV$ . In the high-temperature region, such as the $K_{B}T\approx0.32meV$ , only one maximum value appears in the picture of the dynamic current, and the current reaches the steady-state value soon after reaching this value. In the vicinity of the Kondo temperature, the current oscillates like that of in the high-temperature region. As the temperature begins to drop further, the current starts to oscillate in a new form. For example, at $K_{B}T=0.045meV$, slight oscillation occurs before reaching the steady-state current. The lower the temperature, the larger is the amplitude of the current oscillation, and the more obvious would be the oscillation behavior, such as in the case of $K_{B}T=0.015meV$. It can be seen that in the dynamic transport of the quantum-impurity system, the Kondo effect could be promoted by lowering the temperature to promote the oscillation of the dynamic current and to increase the steady-state current value.\\
\begin{figure}[htbp]
\centering
\includegraphics[width=0.4\textwidth]{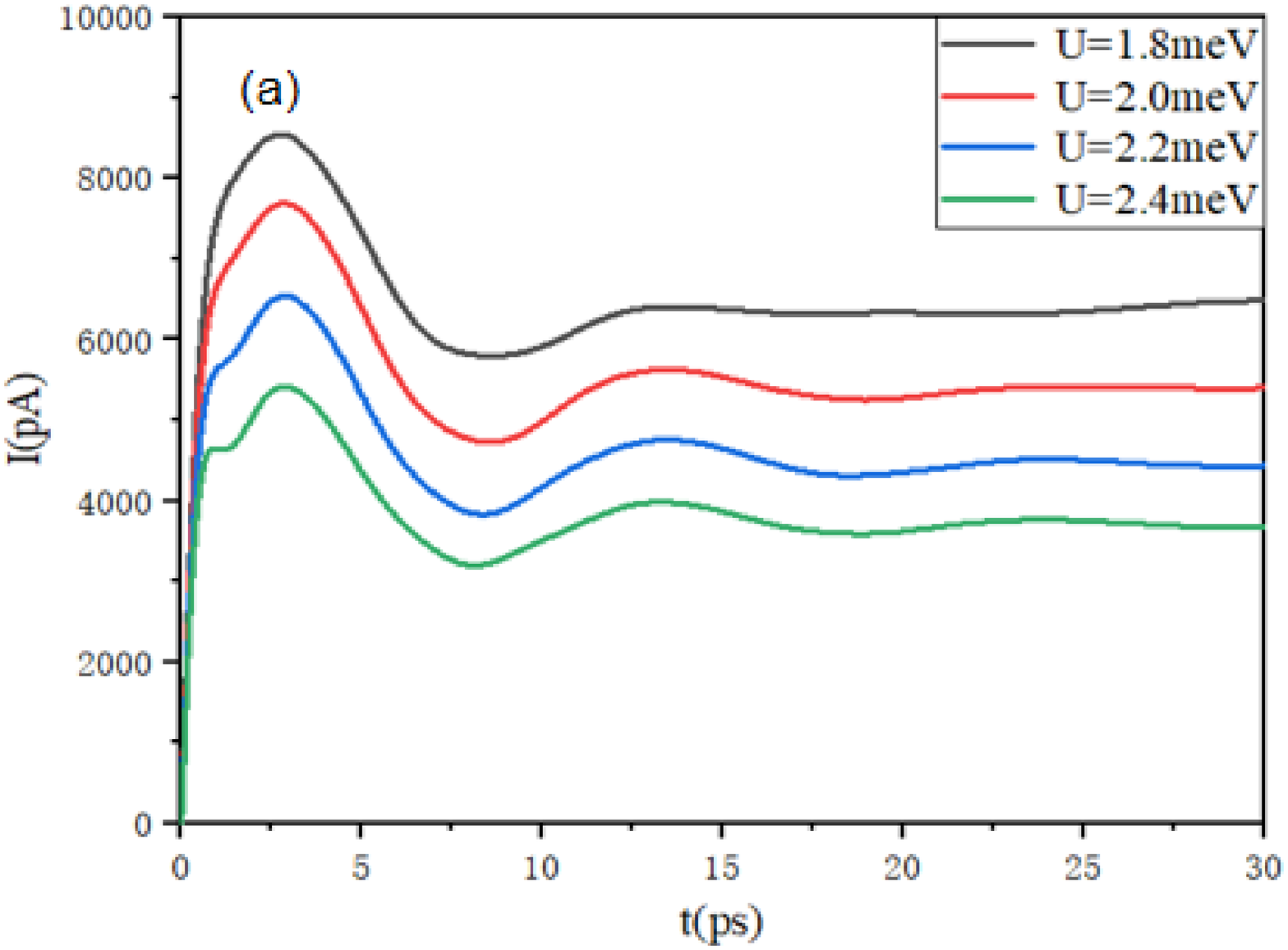}
\includegraphics[width=0.4\textwidth]{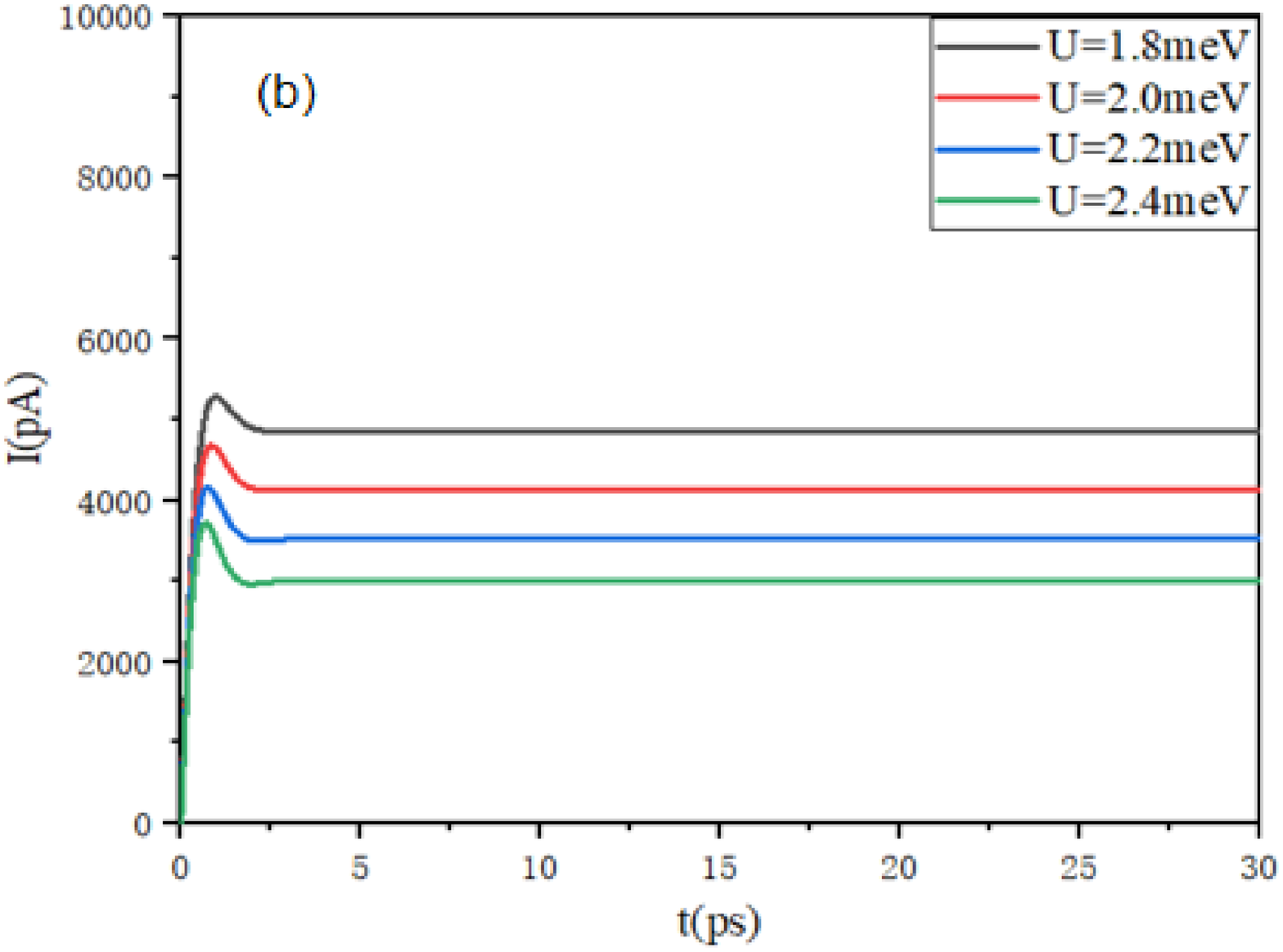}
\caption{The dynamic current of the side-coupled double quantum-impurity system at different e-e interactions U.The parameters adopted are $t_{12}$=0.1meV,$V_{L}=-V_{R}=0.3mV$,$\Delta$=0.2meV,W=2meV,$U_{12}$=0,$\epsilon$$\uparrow$=$\epsilon$$\downarrow$=-1meV.(a)In the Kondo regime,$K_{B}$T=0.015meV.(b)Out of the Kondo regime,$K_{B}$T=0.32meV.}
\end{figure}
Further, the effect of e-e interaction $U$ on the current oscillation was studied. Figure 5 shows the results of the Kondo region $T<T_{K}$ and the non-Kondo region $T>T_{K}$, respectively. The comparison between the two figures shows that the amplitude of the current oscillation decreases with the increase of U, regardless of whether the current is present in the Kondo region or not. This is because the on-site e-e interaction induces the localization of the carrier. Taking a closer look at the Kondo region in figure (a), it can be seen that the amplitude of the current oscillation decreases when U increases, but the frequency remains almost constant. The mechanism can be understood as follows. According to the analytical expression of the Kondo temperature $T_{K}$, Tk increases with the decrease of U. Since the temperature is fixed ($K_{B}T=0.015meV$), a smaller U would result in a larger difference between $T$ and $T_{K}$. Figure 4 indicates that the larger the difference, the stronger would be the current oscillation. Therefore, for smaller U, the Kondo effect of the system would be enhanced, and the current would have a larger amplitude and a higher value of steady-state current. For example, the maximum amplitude of the oscillation increases from 6000 pA of $U=2.4 meV$ to 9000 pA of $U=1.8 meV$, and the steady-state current value increases from 3500 pA of $U=2.4 meV$ to 6400 pA of $U=1.8 meV$.\\
\begin{figure}[htbp]
\centering
\includegraphics[width=0.4\textwidth]{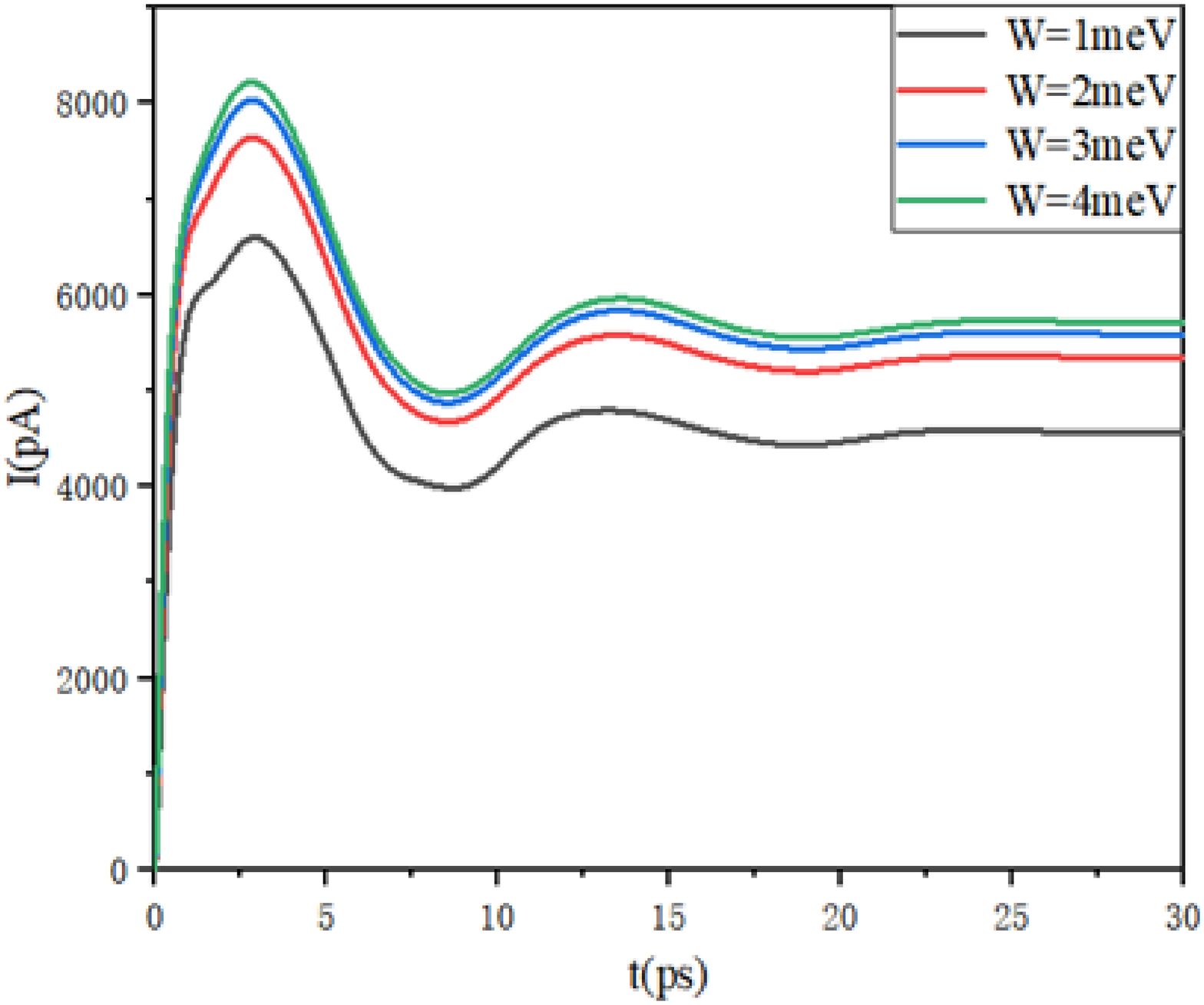}
\caption{The dynamic current of the side-coupled double quantum-impurity system at different band width W.The parameters adopted are $t_{12}$=0.1meV,$V_{L}=-V_{R}=0.3mV$,$K_{B}$T=0.015meV,$\Delta$=0.2meV,U=2meV,$U_{12}$=0,$\epsilon$$\uparrow$=$\epsilon$$\downarrow$=-1meV.}
\end{figure}
Towards the end of this part, we also elucidate the effect of the bandwidth $W$ of the lead on the current oscillation, which is difficult to treat using the nonequilibrium Green's function method under the wideband approximation. The characteristics of the dynamic current corresponding to different bandwidths are shown in Fig. 6. It can be seen that a larger $W$ leads to a relatively larger amplitude of current oscillation and a relatively larger value of the steady-state current. This phenomenon mainly occurs due to the increase in the bandwidth, the accumulation, and consumption of the charge on both sides of the lead increases, resulting in the effect of current enhancement, that is, the bandwidth W-enhanced lead capacitance contribution. In addition, we see that the frequency of oscillation is almost independent of the bandwidth $W$. Therefore, we can roughly assume that the bandwidth of the lead has little effect on the oscillation behavior of the current.\\
\indent
We further ignore the tunneling coupling and study only the pure capacitive coupling.\\
\begin{figure}[htbp]
\centering
\includegraphics[width=0.4\textwidth]{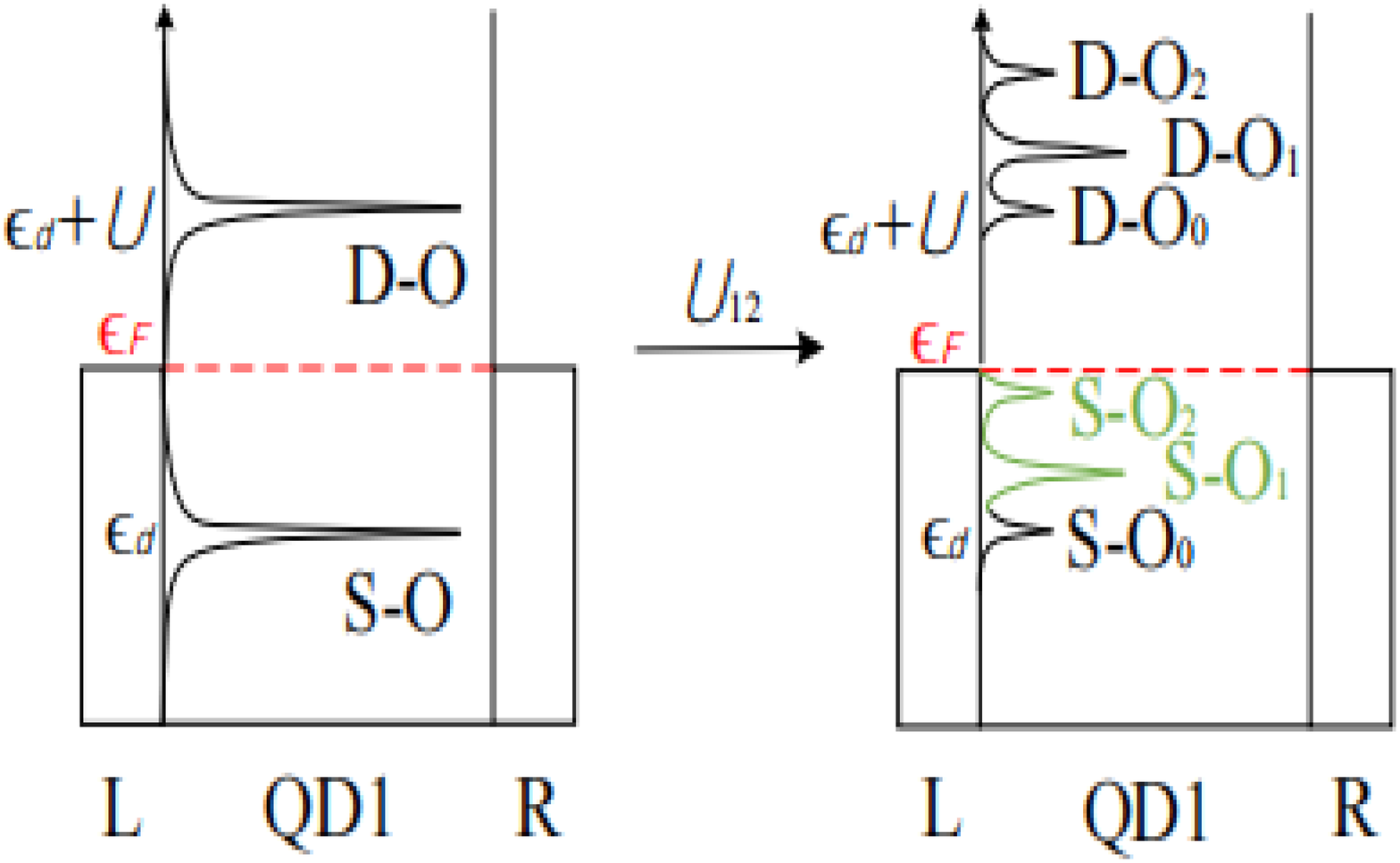}
\caption{Schematic diagram of coulomb coupling configuration of side-coupled double quantum-impurity systems\cite{press2020complete}.}
\end{figure}
\indent
As shown in Fig. 7, when there is no repulsive inter-point coulomb for QD1, there would be two Hubbard bands symmetrically distributed on both sides of the Fermi energy level, representing the singly occupied state below the Fermi energy level ($S-O$ state) and the double occupied state above the Fermi energy level ($D-O$ state), respectively. When $U_{12}$ is applied to the system, the $S-O$ state and the $D-O$ state splits into three quasi-particle states, $SO_{0},SO_{1},SO_{2}$, and $DO_{0},DO_{1},DO_{2}$, respectively. With the increase in $U_{12}$, the $SO_{0}$ and $DO_{0}$ substates remain unchanged, while the $SO_{1}$ and $DO_{1}$ substates deviate from their original position at a rate of 1, and $SO_{2}$ and $DO_{2}$ substates deviate at a rate of 2. When combined with their initial positions, the $SO_{2}$ and $SO_{1}$ substates pass linearly and sequentially through the Fermi energy levels during this process. We find that within the parameter region of $0\le U_{12}\le U$, there exists an interesting interference phenomenon, as shown in the figure below.\\
\begin{figure}[htbp]
\centering
\includegraphics[width=0.4\textwidth]{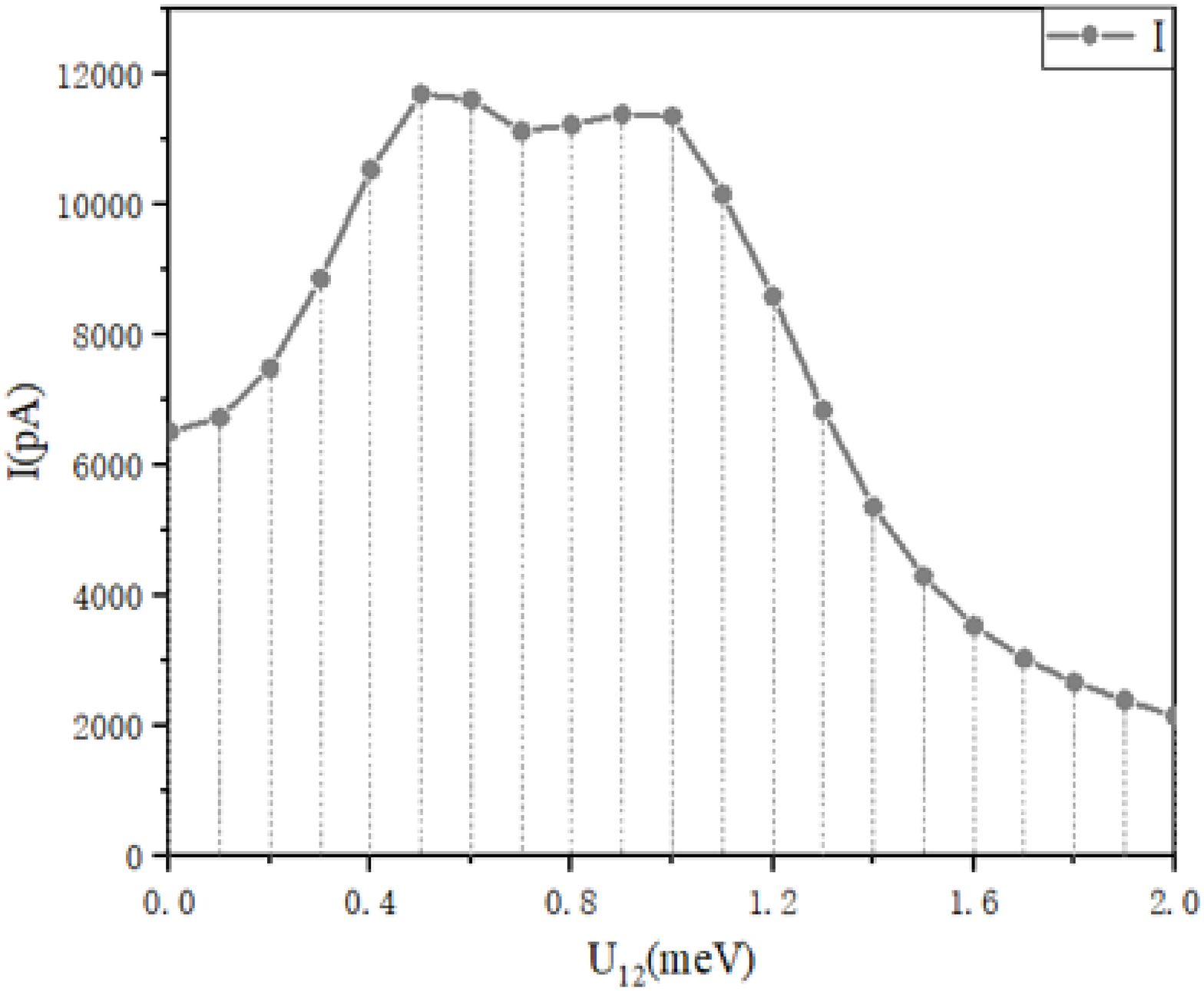}
\caption{The steady state value of current of side-coupled double quantum-impurity systems at different coulomb coupling constant $U_{12}$.The parameters adopted are $t_{12}$=0,$V_{L}=-V_{R}=0.3mV$,$K_{B}$T=0.015meV,$\Delta$=0.2meV,W=2meV,U=2meV,$\epsilon$$\uparrow$=$\epsilon$$\downarrow$=-1meV.}
\end{figure}
\indent
The current steady-state values corresponding to different $U_{12}$ underside coupling double quantum dot systems are described in Fig 8. It can be seen that during $0\le U_{12}\le 0.2meV$, which is in the K-I region, the steady-state current value increases with the increase in $U_{12}$, which is similar to the behavior of a single quantum dot. This is because the $SO_{2}$ first reaches the Fermi level and hence has an interference enhancement effect on the Kondo peak. When $0.2meV<U_{12}<1.2meV$, that is, the system is in the K-F region, the steady-state current presents an“M” type with the increase of $U_{12}$. This is because in this region, with the increase inf $U_{12}$, the up-moving $SO_{1}$ , and $SO_{2}$ state levels and have an impact on the Kondo peak there. When $0.2meV<U_{12}<0.5meV$, the $SO_{2}$ state first reaches the Fermi energy level and brings about the interference enhancement phenomenon of the K-F effect, which indicates that the steady-state current value increases with the increase in $U_{12}$. After that, the $SO_{1}$ state also reaches the Fermi level and influences the Kondo peak together with the $SO_{2}$ state. The strengthening effect of the $SO_{2}$ state and the weakening effect of the $SO_{1}$ state would contradict each other, that is, the linear shape of $0.5meV\le U_{12}\le 1.0meV$ would become a“V” shape. As $U_{12}$ continues to increase, the energy level of the $SO_{2}$ state at $1.0meV<U_{12}<1.2meV$ increases away from the Fermi level, and the $SO_{1}$ state occupies a dominant position; thus, causing the interference suppression phenomenon of the K-F effect. When $1.2meV\le U_{12}\le 2.0meV$, that is, in the K-II region, the steady-state current value decreases $U_{12}$ increases. This is because the $SO_{2}$ state of the Hubbard sub-peak of interference Kondo peak is elevated and situated away from the Fermi level during this time, and only the $SO_{1}$ state has an impact on the system. By comparing the K-I region with the K-II region, it can be seen that only the $SO_{2}$ state and the $SO_{1}$ state possess different interference effects on the Kondo summit.\\
\begin{figure}[htbp]
\centering
\includegraphics[width=0.4\textwidth]{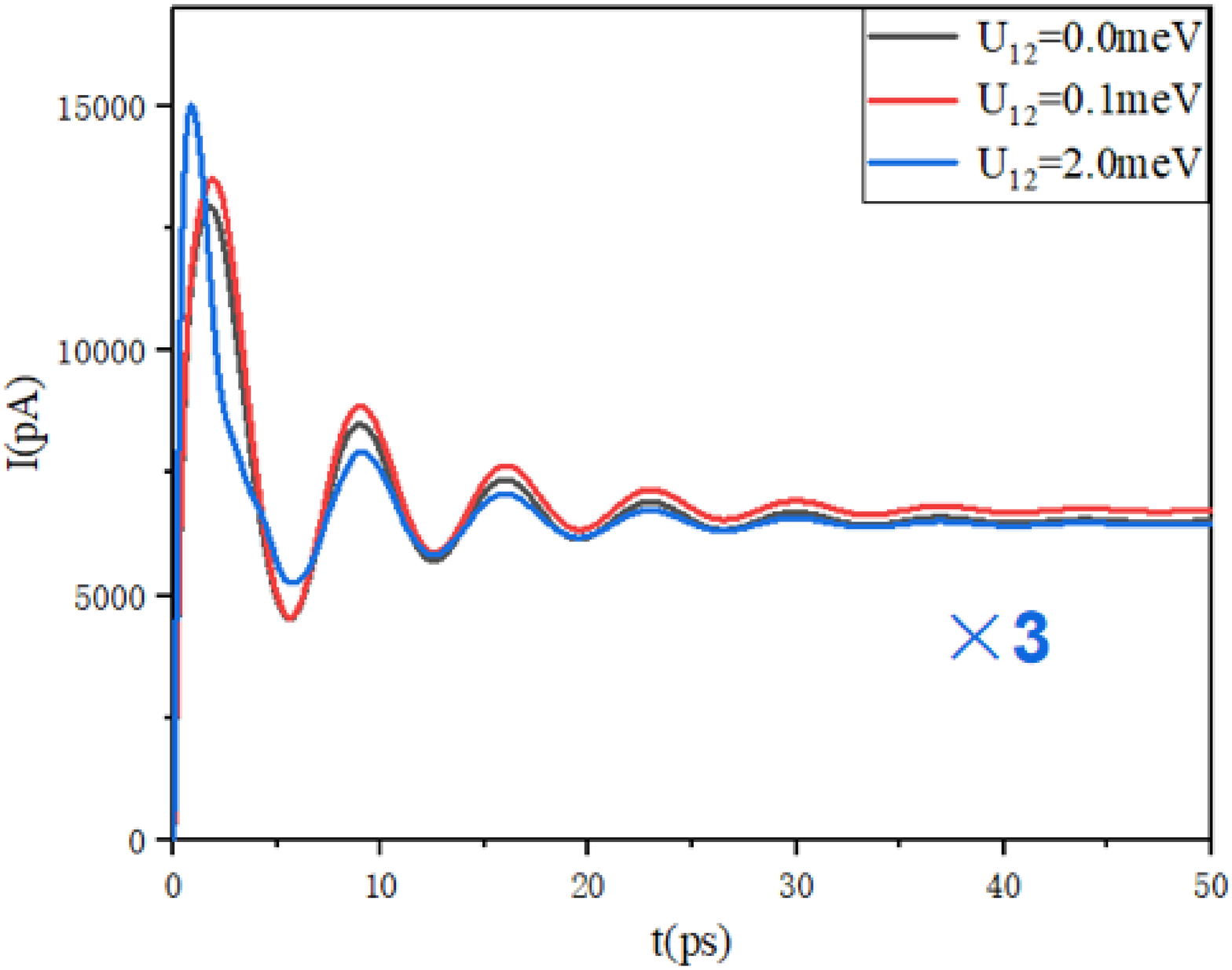}
\caption{The current of side-coupled double quantum-impurity systems at different coulomb coupling constant $U_{12}$.The parameters adopted are $t_{12}$=0,$V_{L}=-V_{R}=0.3mV$,$K_{B}$T=0.015meV,$\Delta$=0.2meV,W=2meV,U=2meV,$\epsilon$$\uparrow$=$\epsilon$$\downarrow$=-1meV.}
\end{figure}
Fig. 9 shows the dynamic current transport images of the single quantum dot, K-I region, and the K-II region respectively. It is seen that the dynamic transport current image within the K-I region ($U_{12} = 0.1meV$) is roughly the same as that of the linear shape in the single quantum dot, and the dynamic transport current image in the K-II region is 1/3 the value of the single quantum dot at large $U_{12} $ ($U_{12} = 2meV$). This is because the coupling strength is weak, and the system behaves like a single quantum dot in the K-I region. However, in the K-II region, when the Coulomb coupling strength is very high, the $SO_{2}$ state is far away from the Fermi energy level, and the $SO_{1}$ state would have a major influence on the Kondo peak. The lines in this part are thus similar to those in the K-I region but consist of only a third of the value. 
\section{\bf{Conclusion}}
To summarize, we studied the time-dependent dynamic current transport properties of the side-coupled double quantum-impurity system based on the HEOM method under different coupling conditions. The transport current would behave like a single quantum dot when the coupling strength is low during both tunneling and capacitive coupling.\\
\indent
Only for the tunneling transition in the Kondo region, the dynamic current oscillates due to the temporal coherence of the electron tunneling device. The oscillation frequency of the transport current is strongly dependent on the bias voltage applied by the lead and is insensitive to the change in values of $T$, $U$ and $W$. The amplitude of the current oscillation is in positive correlation with the electrode bandwidth $W$ and in negative correlation with the e-e interaction $U$. With the increase of the coupling $t_{12}$ between impurities, the ground state of the system changes from a Kondo singlet of single impurity to a spin-singlet of the numerical calculation, such that the system would experience three-parameter regions: the Kondo-I region, Kondo-Fano region and Kondo-II region with the change of $U_{12}$. When the $U_{12}$ value is very small, the current in the K-I region behaves like that in a single quantum-impurity system, and only the $SO_{2}$ states would have a very weak interference enhancement effect on the Kondo peak. With the increase of $U_{12}$, two sub-summits present below the Fermi energy level are elevated and are passed through the Fermi level in turn. They produce fano interference on the Kondo peak, forming the K-F effect, which describes the phenomenon of interference enhancement, interference competition, and interference suppression. As the $U_{12}$ value continues to increase, the Hubbard sub-peak, which interferes with the Kondo peak, continues to elevate, such that it lies far away from the Fermi level. At this point, the transport current in the K-II region shows qualitative consistency with that in the K-I region, with a 1/3 quantitative relationship.\\
\indent
Our next plan of work is to simultaneously change the tunneling transition coupling $t_{12}$ and the coulomb interaction parameter $U_{12}$ to observe their effects on the dynamic transport current of the side-coupled double quantum-impurity system.
\section{acknowledgement}
The support from the Natural Science Foundation of China {(Grant Nos. 11774418 and 11374363)} is gratefully appreciated. Computational resources have been provided by the Physical Laboratory of High Performance Computing at Renmin University of China.
\bibliography{aa}

\begin{thebibliography}{36}%
\makeatletter
\providecommand \@ifxundefined [1]{%
 \@ifx{#1\undefined}
}%
\providecommand \@ifnum [1]{%
 \ifnum #1\expandafter \@firstoftwo
 \else \expandafter \@secondoftwo
 \fi
}%
\providecommand \@ifx [1]{%
 \ifx #1\expandafter \@firstoftwo
 \else \expandafter \@secondoftwo
 \fi
}%
\providecommand \natexlab [1]{#1}%
\providecommand \enquote  [1]{``#1''}%
\providecommand \bibnamefont  [1]{#1}%
\providecommand \bibfnamefont [1]{#1}%
\providecommand \citenamefont [1]{#1}%
\providecommand \href@noop [0]{\@secondoftwo}%
\providecommand \href [0]{\begingroup \@sanitize@url \@href}%
\providecommand \@href[1]{\@@startlink{#1}\@@href}%
\providecommand \@@href[1]{\endgroup#1\@@endlink}%
\providecommand \@sanitize@url [0]{\catcode `\\12\catcode `\$12\catcode
  `\&12\catcode `\#12\catcode `\^12\catcode `\_12\catcode `\%12\relax}%
\providecommand \@@startlink[1]{}%
\providecommand \@@endlink[0]{}%
\providecommand \url  [0]{\begingroup\@sanitize@url \@url }%
\providecommand \@url [1]{\endgroup\@href {#1}{\urlprefix }}%
\providecommand \urlprefix  [0]{URL }%
\providecommand \Eprint [0]{\href }%
\providecommand \doibase [0]{http://dx.doi.org/}%
\providecommand \selectlanguage [0]{\@gobble}%
\providecommand \bibinfo  [0]{\@secondoftwo}%
\providecommand \bibfield  [0]{\@secondoftwo}%
\providecommand \translation [1]{[#1]}%
\providecommand \BibitemOpen [0]{}%
\providecommand \bibitemStop [0]{}%
\providecommand \bibitemNoStop [0]{.\EOS\space}%
\providecommand \EOS [0]{\spacefactor3000\relax}%
\providecommand \BibitemShut  [1]{\csname bibitem#1\endcsname}%
\let\auto@bib@innerbib\@empty
\bibitem [{\citenamefont {Zhao}\ \emph {et~al.}(2019)\citenamefont {Zhao},
  \citenamefont {Min},\ and\ \citenamefont {Huang}}]{1}%
  \BibitemOpen
  \bibfield  {author} {\bibinfo {author} {\bibfnamefont {Z.}~\bibnamefont
  {Zhao}}, \bibinfo {author} {\bibfnamefont {Y.}~\bibnamefont {Min}}, \ and\
  \bibinfo {author} {\bibfnamefont {Y.}~\bibnamefont {Huang}},\ }\href
  {\doibase https://doi.org/10.1016/j.physe.2019.113589} {\bibfield  {journal}
  {\bibinfo  {journal} {Physica E: Low-dimensional Systems and Nanostructures}\
  }\textbf {\bibinfo {volume} {114}},\ \bibinfo {pages} {113589} (\bibinfo
  {year} {2019})}\BibitemShut {NoStop}%
\bibitem [{\citenamefont {Jin}\ \emph {et~al.}(2018)\citenamefont {Jin},
  \citenamefont {Wang}, \citenamefont {Zhou}, \citenamefont {Zhang},\ and\
  \citenamefont {Yan}}]{2}%
  \BibitemOpen
  \bibfield  {author} {\bibinfo {author} {\bibfnamefont {J.}~\bibnamefont
  {Jin}}, \bibinfo {author} {\bibfnamefont {S.}~\bibnamefont {Wang}}, \bibinfo
  {author} {\bibfnamefont {J.}~\bibnamefont {Zhou}}, \bibinfo {author}
  {\bibfnamefont {W.-M.}\ \bibnamefont {Zhang}}, \ and\ \bibinfo {author}
  {\bibfnamefont {Y.}~\bibnamefont {Yan}},\ }\href {\doibase
  10.1088/1367-2630/aab5cb} {\bibfield  {journal} {\bibinfo  {journal} {New
  Journal of Physics}\ }\textbf {\bibinfo {volume} {20}},\ \bibinfo {pages}
  {043043} (\bibinfo {year} {2018})}\BibitemShut {NoStop}%
\bibitem [{\citenamefont {Yang}\ \emph {et~al.}(2021)\citenamefont {Yang},
  \citenamefont {Di}, \citenamefont {Wang}, \citenamefont {Wang},\ and\
  \citenamefont {ai~Yang}}]{3}%
  \BibitemOpen
  \bibfield  {author} {\bibinfo {author} {\bibfnamefont {K.-H.}\ \bibnamefont
  {Yang}}, \bibinfo {author} {\bibfnamefont {H.-Y.}\ \bibnamefont {Di}},
  \bibinfo {author} {\bibfnamefont {H.-Y.}\ \bibnamefont {Wang}}, \bibinfo
  {author} {\bibfnamefont {X.}~\bibnamefont {Wang}}, \ and\ \bibinfo {author}
  {\bibfnamefont {A.}~\bibnamefont {ai~Yang}},\ }\href {\doibase
  https://doi.org/10.1016/j.physleta.2020.127095} {\bibfield  {journal}
  {\bibinfo  {journal} {Physics Letters A}\ }\textbf {\bibinfo {volume}
  {389}},\ \bibinfo {pages} {127095} (\bibinfo {year} {2021})}\BibitemShut
  {NoStop}%
\bibitem [{\citenamefont {Sun}\ \emph {et~al.}(2020)\citenamefont {Sun},
  \citenamefont {Wang}, \citenamefont {Wei},\ and\ \citenamefont {Yan}}]{4}%
  \BibitemOpen
  \bibfield  {author} {\bibinfo {author} {\bibfnamefont {F.-L.}\ \bibnamefont
  {Sun}}, \bibinfo {author} {\bibfnamefont {Y.-D.}\ \bibnamefont {Wang}},
  \bibinfo {author} {\bibfnamefont {J.-H.}\ \bibnamefont {Wei}}, \ and\
  \bibinfo {author} {\bibfnamefont {Y.-J.}\ \bibnamefont {Yan}},\ }\href
  {\doibase 10.1088/1674-1056/ab8ac1} {\bibfield  {journal} {\bibinfo
  {journal} {Chinese Physics B}\ }\textbf {\bibinfo {volume} {29}},\ \bibinfo
  {pages} {067204} (\bibinfo {year} {2020})}\BibitemShut {NoStop}%
\bibitem [{\citenamefont {Aguado}\ and\ \citenamefont {Langreth}(2000)}]{5}%
  \BibitemOpen
  \bibfield  {author} {\bibinfo {author} {\bibfnamefont {R.}~\bibnamefont
  {Aguado}}\ and\ \bibinfo {author} {\bibfnamefont {D.~C.}\ \bibnamefont
  {Langreth}},\ }\href {\doibase 10.1103/physrevlett.85.1946} {\bibfield
  {journal} {\bibinfo  {journal} {Physical Review Letters}\ }\textbf {\bibinfo
  {volume} {85}},\ \bibinfo {pages} {1946–1949} (\bibinfo {year}
  {2000})}\BibitemShut {NoStop}%
\bibitem [{\citenamefont {Izumida}\ and\ \citenamefont {Sakai}(2000)}]{6}%
  \BibitemOpen
  \bibfield  {author} {\bibinfo {author} {\bibfnamefont {W.}~\bibnamefont
  {Izumida}}\ and\ \bibinfo {author} {\bibfnamefont {O.}~\bibnamefont
  {Sakai}},\ }\href {\doibase 10.1103/PhysRevB.62.10260} {\bibfield  {journal}
  {\bibinfo  {journal} {Phys. Rev. B}\ }\textbf {\bibinfo {volume} {62}},\
  \bibinfo {pages} {10260} (\bibinfo {year} {2000})}\BibitemShut {NoStop}%
\bibitem [{\citenamefont {Wang}(2011)}]{7}%
  \BibitemOpen
  \bibfield  {author} {\bibinfo {author} {\bibfnamefont {W.-z.}\ \bibnamefont
  {Wang}},\ }\href {\doibase 10.1103/PhysRevB.83.075314} {\bibfield  {journal}
  {\bibinfo  {journal} {Phys. Rev. B}\ }\textbf {\bibinfo {volume} {83}},\
  \bibinfo {pages} {075314} (\bibinfo {year} {2011})}\BibitemShut {NoStop}%
\bibitem [{\citenamefont {Galpin}\ \emph {et~al.}(2005)\citenamefont {Galpin},
  \citenamefont {Logan},\ and\ \citenamefont {Krishnamurthy}}]{8}%
  \BibitemOpen
  \bibfield  {author} {\bibinfo {author} {\bibfnamefont {M.~R.}\ \bibnamefont
  {Galpin}}, \bibinfo {author} {\bibfnamefont {D.~E.}\ \bibnamefont {Logan}}, \
  and\ \bibinfo {author} {\bibfnamefont {H.~R.}\ \bibnamefont
  {Krishnamurthy}},\ }\href {\doibase 10.1103/PhysRevLett.94.186406} {\bibfield
   {journal} {\bibinfo  {journal} {Phys. Rev. Lett.}\ }\textbf {\bibinfo
  {volume} {94}},\ \bibinfo {pages} {186406} (\bibinfo {year}
  {2005})}\BibitemShut {NoStop}%
\bibitem [{\citenamefont {Trocha}\ and\ \citenamefont
  {Barna\ifmmode~\acute{s}\else \'{s}\fi{}}(2012)}]{9}%
  \BibitemOpen
  \bibfield  {author} {\bibinfo {author} {\bibfnamefont {P.}~\bibnamefont
  {Trocha}}\ and\ \bibinfo {author} {\bibfnamefont {J.}~\bibnamefont
  {Barna\ifmmode~\acute{s}\else \'{s}\fi{}}},\ }\href {\doibase
  10.1103/PhysRevB.85.085408} {\bibfield  {journal} {\bibinfo  {journal} {Phys.
  Rev. B}\ }\textbf {\bibinfo {volume} {85}},\ \bibinfo {pages} {085408}
  (\bibinfo {year} {2012})}\BibitemShut {NoStop}%
\bibitem [{\citenamefont {Juergens}\ \emph {et~al.}(2013)\citenamefont
  {Juergens}, \citenamefont {Haupt}, \citenamefont {Moskalets},\ and\
  \citenamefont {Splettstoesser}}]{10}%
  \BibitemOpen
  \bibfield  {author} {\bibinfo {author} {\bibfnamefont {S.}~\bibnamefont
  {Juergens}}, \bibinfo {author} {\bibfnamefont {F.}~\bibnamefont {Haupt}},
  \bibinfo {author} {\bibfnamefont {M.}~\bibnamefont {Moskalets}}, \ and\
  \bibinfo {author} {\bibfnamefont {J.}~\bibnamefont {Splettstoesser}},\ }\href
  {\doibase 10.1103/PhysRevB.87.245423} {\bibfield  {journal} {\bibinfo
  {journal} {Phys. Rev. B}\ }\textbf {\bibinfo {volume} {87}},\ \bibinfo
  {pages} {245423} (\bibinfo {year} {2013})}\BibitemShut {NoStop}%
\bibitem [{\citenamefont {Li}\ \emph {et~al.}(2018)\citenamefont {Li},
  \citenamefont {Chen}, \citenamefont {Muga},\ and\ \citenamefont
  {Sherman}}]{11}%
  \BibitemOpen
  \bibfield  {author} {\bibinfo {author} {\bibfnamefont {Y.-C.}\ \bibnamefont
  {Li}}, \bibinfo {author} {\bibfnamefont {X.}~\bibnamefont {Chen}}, \bibinfo
  {author} {\bibfnamefont {J.~G.}\ \bibnamefont {Muga}}, \ and\ \bibinfo
  {author} {\bibfnamefont {E.~Y.}\ \bibnamefont {Sherman}},\ }\href {\doibase
  10.1088/1367-2630/aaedd9} {\bibfield  {journal} {\bibinfo  {journal} {New
  Journal of Physics}\ }\textbf {\bibinfo {volume} {20}},\ \bibinfo {pages}
  {113029} (\bibinfo {year} {2018})}\BibitemShut {NoStop}%
\bibitem [{\citenamefont {Karwat}\ and\ \citenamefont
  {Machnikowski}(2015)}]{12}%
  \BibitemOpen
  \bibfield  {author} {\bibinfo {author} {\bibfnamefont {P.}~\bibnamefont
  {Karwat}}\ and\ \bibinfo {author} {\bibfnamefont {P.}~\bibnamefont
  {Machnikowski}},\ }\href {\doibase 10.1103/PhysRevB.91.125428} {\bibfield
  {journal} {\bibinfo  {journal} {Phys. Rev. B}\ }\textbf {\bibinfo {volume}
  {91}},\ \bibinfo {pages} {125428} (\bibinfo {year} {2015})}\BibitemShut
  {NoStop}%
\bibitem [{\citenamefont {Das}\ \emph {et~al.}(2008)\citenamefont {Das},
  \citenamefont {Agarwal},\ and\ \citenamefont {Scully}}]{13}%
  \BibitemOpen
  \bibfield  {author} {\bibinfo {author} {\bibfnamefont {S.}~\bibnamefont
  {Das}}, \bibinfo {author} {\bibfnamefont {G.~S.}\ \bibnamefont {Agarwal}}, \
  and\ \bibinfo {author} {\bibfnamefont {M.~O.}\ \bibnamefont {Scully}},\
  }\href {\doibase 10.1103/PhysRevLett.101.153601} {\bibfield  {journal}
  {\bibinfo  {journal} {Phys. Rev. Lett.}\ }\textbf {\bibinfo {volume} {101}},\
  \bibinfo {pages} {153601} (\bibinfo {year} {2008})}\BibitemShut {NoStop}%
\bibitem [{\citenamefont {Landauer}(1957)}]{14}%
  \BibitemOpen
  \bibfield  {author} {\bibinfo {author} {\bibfnamefont {R.}~\bibnamefont
  {Landauer}},\ }\href {\doibase 10.1147/rd.13.0223} {\bibfield  {journal}
  {\bibinfo  {journal} {IBM Journal of Research and Development}\ }\textbf
  {\bibinfo {volume} {1}},\ \bibinfo {pages} {223} (\bibinfo {year}
  {1957})}\BibitemShut {NoStop}%
\bibitem [{\citenamefont {B\"uttiker}(1986)}]{15}%
  \BibitemOpen
  \bibfield  {author} {\bibinfo {author} {\bibfnamefont {M.}~\bibnamefont
  {B\"uttiker}},\ }\href {\doibase 10.1103/PhysRevLett.57.1761} {\bibfield
  {journal} {\bibinfo  {journal} {Phys. Rev. Lett.}\ }\textbf {\bibinfo
  {volume} {57}},\ \bibinfo {pages} {1761} (\bibinfo {year}
  {1986})}\BibitemShut {NoStop}%
\bibitem [{\citenamefont {Wingreen}\ \emph {et~al.}(1993)\citenamefont
  {Wingreen}, \citenamefont {Jauho},\ and\ \citenamefont {Meir}}]{16}%
  \BibitemOpen
  \bibfield  {author} {\bibinfo {author} {\bibfnamefont {N.~S.}\ \bibnamefont
  {Wingreen}}, \bibinfo {author} {\bibfnamefont {A.-P.}\ \bibnamefont {Jauho}},
  \ and\ \bibinfo {author} {\bibfnamefont {Y.}~\bibnamefont {Meir}},\ }\href
  {\doibase 10.1103/PhysRevB.48.8487} {\bibfield  {journal} {\bibinfo
  {journal} {Phys. Rev. B}\ }\textbf {\bibinfo {volume} {48}},\ \bibinfo
  {pages} {8487} (\bibinfo {year} {1993})}\BibitemShut {NoStop}%
\bibitem [{\citenamefont {Jauho}\ \emph {et~al.}(1994)\citenamefont {Jauho},
  \citenamefont {Wingreen},\ and\ \citenamefont {Meir}}]{17}%
  \BibitemOpen
  \bibfield  {author} {\bibinfo {author} {\bibfnamefont {A.-P.}\ \bibnamefont
  {Jauho}}, \bibinfo {author} {\bibfnamefont {N.~S.}\ \bibnamefont {Wingreen}},
  \ and\ \bibinfo {author} {\bibfnamefont {Y.}~\bibnamefont {Meir}},\ }\href
  {\doibase 10.1103/PhysRevB.50.5528} {\bibfield  {journal} {\bibinfo
  {journal} {Phys. Rev. B}\ }\textbf {\bibinfo {volume} {50}},\ \bibinfo
  {pages} {5528} (\bibinfo {year} {1994})}\BibitemShut {NoStop}%
\bibitem [{\citenamefont {Zhu}\ \emph {et~al.}(2005)\citenamefont {Zhu},
  \citenamefont {Maciejko}, \citenamefont {Ji}, \citenamefont {Guo},\ and\
  \citenamefont {Wang}}]{18}%
  \BibitemOpen
  \bibfield  {author} {\bibinfo {author} {\bibfnamefont {Y.}~\bibnamefont
  {Zhu}}, \bibinfo {author} {\bibfnamefont {J.}~\bibnamefont {Maciejko}},
  \bibinfo {author} {\bibfnamefont {T.}~\bibnamefont {Ji}}, \bibinfo {author}
  {\bibfnamefont {H.}~\bibnamefont {Guo}}, \ and\ \bibinfo {author}
  {\bibfnamefont {J.}~\bibnamefont {Wang}},\ }\href {\doibase
  10.1103/PhysRevB.71.075317} {\bibfield  {journal} {\bibinfo  {journal} {Phys.
  Rev. B}\ }\textbf {\bibinfo {volume} {71}},\ \bibinfo {pages} {075317}
  (\bibinfo {year} {2005})}\BibitemShut {NoStop}%
\bibitem [{\citenamefont {Maciejko}\ \emph {et~al.}(2006)\citenamefont
  {Maciejko}, \citenamefont {Wang},\ and\ \citenamefont {Guo}}]{19}%
  \BibitemOpen
  \bibfield  {author} {\bibinfo {author} {\bibfnamefont {J.}~\bibnamefont
  {Maciejko}}, \bibinfo {author} {\bibfnamefont {J.}~\bibnamefont {Wang}}, \
  and\ \bibinfo {author} {\bibfnamefont {H.}~\bibnamefont {Guo}},\ }\href
  {\doibase 10.1103/physrevb.74.085324} {\bibfield  {journal} {\bibinfo
  {journal} {Physical Review B}\ }\textbf {\bibinfo {volume} {74}} (\bibinfo
  {year} {2006}),\ 10.1103/physrevb.74.085324}\BibitemShut {NoStop}%
\bibitem [{\citenamefont {Cazalilla}\ and\ \citenamefont {Marston}(2002)}]{20}%
  \BibitemOpen
  \bibfield  {author} {\bibinfo {author} {\bibfnamefont {M.~A.}\ \bibnamefont
  {Cazalilla}}\ and\ \bibinfo {author} {\bibfnamefont {J.~B.}\ \bibnamefont
  {Marston}},\ }\href {\doibase 10.1103/PhysRevLett.88.256403} {\bibfield
  {journal} {\bibinfo  {journal} {Phys. Rev. Lett.}\ }\textbf {\bibinfo
  {volume} {88}},\ \bibinfo {pages} {256403} (\bibinfo {year}
  {2002})}\BibitemShut {NoStop}%
\bibitem [{\citenamefont {Schmitteckert}(2004)}]{21}%
  \BibitemOpen
  \bibfield  {author} {\bibinfo {author} {\bibfnamefont {P.}~\bibnamefont
  {Schmitteckert}},\ }\href {\doibase 10.1103/physrevb.70.121302} {\bibfield
  {journal} {\bibinfo  {journal} {Physical Review B}\ }\textbf {\bibinfo
  {volume} {70}} (\bibinfo {year} {2004}),\
  10.1103/physrevb.70.121302}\BibitemShut {NoStop}%
\bibitem [{\citenamefont {Heidrich-Meisner}\ \emph {et~al.}(2009)\citenamefont
  {Heidrich-Meisner}, \citenamefont {Feiguin},\ and\ \citenamefont
  {Dagotto}}]{22}%
  \BibitemOpen
  \bibfield  {author} {\bibinfo {author} {\bibfnamefont {F.}~\bibnamefont
  {Heidrich-Meisner}}, \bibinfo {author} {\bibfnamefont {A.~E.}\ \bibnamefont
  {Feiguin}}, \ and\ \bibinfo {author} {\bibfnamefont {E.}~\bibnamefont
  {Dagotto}},\ }\href {\doibase 10.1103/PhysRevB.79.235336} {\bibfield
  {journal} {\bibinfo  {journal} {Phys. Rev. B}\ }\textbf {\bibinfo {volume}
  {79}},\ \bibinfo {pages} {235336} (\bibinfo {year} {2009})}\BibitemShut
  {NoStop}%
\bibitem [{\citenamefont {Anders}\ and\ \citenamefont {Schiller}(2005)}]{23}%
  \BibitemOpen
  \bibfield  {author} {\bibinfo {author} {\bibfnamefont {F.~B.}\ \bibnamefont
  {Anders}}\ and\ \bibinfo {author} {\bibfnamefont {A.}~\bibnamefont
  {Schiller}},\ }\href {\doibase 10.1103/PhysRevLett.95.196801} {\bibfield
  {journal} {\bibinfo  {journal} {Phys. Rev. Lett.}\ }\textbf {\bibinfo
  {volume} {95}},\ \bibinfo {pages} {196801} (\bibinfo {year}
  {2005})}\BibitemShut {NoStop}%
\bibitem [{\citenamefont {Anders}\ and\ \citenamefont {Schiller}(2006)}]{24}%
  \BibitemOpen
  \bibfield  {author} {\bibinfo {author} {\bibfnamefont {F.~B.}\ \bibnamefont
  {Anders}}\ and\ \bibinfo {author} {\bibfnamefont {A.}~\bibnamefont
  {Schiller}},\ }\href {\doibase 10.1103/PhysRevB.74.245113} {\bibfield
  {journal} {\bibinfo  {journal} {Phys. Rev. B}\ }\textbf {\bibinfo {volume}
  {74}},\ \bibinfo {pages} {245113} (\bibinfo {year} {2006})}\BibitemShut
  {NoStop}%
\bibitem [{\citenamefont {Cheng}\ \emph {et~al.}(2015)\citenamefont {Cheng},
  \citenamefont {Hou}, \citenamefont {Wang}, \citenamefont {Li}, \citenamefont
  {Wei},\ and\ \citenamefont {Yan}}]{25}%
  \BibitemOpen
  \bibfield  {author} {\bibinfo {author} {\bibfnamefont {Y.}~\bibnamefont
  {Cheng}}, \bibinfo {author} {\bibfnamefont {W.}~\bibnamefont {Hou}}, \bibinfo
  {author} {\bibfnamefont {Y.}~\bibnamefont {Wang}}, \bibinfo {author}
  {\bibfnamefont {Z.}~\bibnamefont {Li}}, \bibinfo {author} {\bibfnamefont
  {J.}~\bibnamefont {Wei}}, \ and\ \bibinfo {author} {\bibfnamefont
  {Y.}~\bibnamefont {Yan}},\ }\href {\doibase 10.1088/1367-2630/17/3/033009}
  {\bibfield  {journal} {\bibinfo  {journal} {New Journal of Physics}\ }\textbf
  {\bibinfo {volume} {17}},\ \bibinfo {pages} {033009} (\bibinfo {year}
  {2015})}\BibitemShut {NoStop}%
\bibitem [{\citenamefont {Jin}\ \emph {et~al.}(2008)\citenamefont {Jin},
  \citenamefont {Zheng},\ and\ \citenamefont {Yan}}]{26}%
  \BibitemOpen
  \bibfield  {author} {\bibinfo {author} {\bibfnamefont {J.}~\bibnamefont
  {Jin}}, \bibinfo {author} {\bibfnamefont {X.}~\bibnamefont {Zheng}}, \ and\
  \bibinfo {author} {\bibfnamefont {Y.}~\bibnamefont {Yan}},\ }\href {\doibase
  10.1063/1.2938087} {\bibfield  {journal} {\bibinfo  {journal} {The Journal of
  Chemical Physics}\ }\textbf {\bibinfo {volume} {128}},\ \bibinfo {pages}
  {234703} (\bibinfo {year} {2008})}\BibitemShut {NoStop}%
\bibitem [{\citenamefont {Zheng}\ \emph {et~al.}(2008)\citenamefont {Zheng},
  \citenamefont {Jin},\ and\ \citenamefont {Yan}}]{27}%
  \BibitemOpen
  \bibfield  {author} {\bibinfo {author} {\bibfnamefont {X.}~\bibnamefont
  {Zheng}}, \bibinfo {author} {\bibfnamefont {J.}~\bibnamefont {Jin}}, \ and\
  \bibinfo {author} {\bibfnamefont {Y.}~\bibnamefont {Yan}},\ }\href {\doibase
  10.1088/1367-2630/10/9/093016} {\bibfield  {journal} {\bibinfo  {journal}
  {New Journal of Physics}\ }\textbf {\bibinfo {volume} {10}},\ \bibinfo
  {pages} {093016} (\bibinfo {year} {2008})}\BibitemShut {NoStop}%
\bibitem [{\citenamefont {Zheng}\ \emph {et~al.}(2009)\citenamefont {Zheng},
  \citenamefont {Luo}, \citenamefont {Jin},\ and\ \citenamefont {Yan}}]{28}%
  \BibitemOpen
  \bibfield  {author} {\bibinfo {author} {\bibfnamefont {X.}~\bibnamefont
  {Zheng}}, \bibinfo {author} {\bibfnamefont {J.}~\bibnamefont {Luo}}, \bibinfo
  {author} {\bibfnamefont {J.}~\bibnamefont {Jin}}, \ and\ \bibinfo {author}
  {\bibfnamefont {Y.}~\bibnamefont {Yan}},\ }\href {\doibase 10.1063/1.3095424}
  {\bibfield  {journal} {\bibinfo  {journal} {The Journal of Chemical Physics}\
  }\textbf {\bibinfo {volume} {130}},\ \bibinfo {pages} {124508} (\bibinfo
  {year} {2009})},\ \Eprint
  {http://arxiv.org/abs/https://doi.org/10.1063/1.3095424}
  {https://doi.org/10.1063/1.3095424} \BibitemShut {NoStop}%
\bibitem [{\citenamefont {Li}\ \emph {et~al.}(2012)\citenamefont {Li},
  \citenamefont {Tong}, \citenamefont {Zheng}, \citenamefont {Hou},
  \citenamefont {Wei}, \citenamefont {Hu},\ and\ \citenamefont {Yan}}]{29}%
  \BibitemOpen
  \bibfield  {author} {\bibinfo {author} {\bibfnamefont {Z.}~\bibnamefont
  {Li}}, \bibinfo {author} {\bibfnamefont {N.}~\bibnamefont {Tong}}, \bibinfo
  {author} {\bibfnamefont {X.}~\bibnamefont {Zheng}}, \bibinfo {author}
  {\bibfnamefont {D.}~\bibnamefont {Hou}}, \bibinfo {author} {\bibfnamefont
  {J.}~\bibnamefont {Wei}}, \bibinfo {author} {\bibfnamefont {J.}~\bibnamefont
  {Hu}}, \ and\ \bibinfo {author} {\bibfnamefont {Y.}~\bibnamefont {Yan}},\
  }\href {\doibase 10.1103/PhysRevLett.109.266403} {\bibfield  {journal}
  {\bibinfo  {journal} {Phys. Rev. Lett.}\ }\textbf {\bibinfo {volume} {109}},\
  \bibinfo {pages} {266403} (\bibinfo {year} {2012})}\BibitemShut {NoStop}%
\bibitem [{\citenamefont {Zheng}\ \emph {et~al.}(2013)\citenamefont {Zheng},
  \citenamefont {Yan},\ and\ \citenamefont {Di~Ventra}}]{30}%
  \BibitemOpen
  \bibfield  {author} {\bibinfo {author} {\bibfnamefont {X.}~\bibnamefont
  {Zheng}}, \bibinfo {author} {\bibfnamefont {Y.}~\bibnamefont {Yan}}, \ and\
  \bibinfo {author} {\bibfnamefont {M.}~\bibnamefont {Di~Ventra}},\ }\href
  {\doibase 10.1103/PhysRevLett.111.086601} {\bibfield  {journal} {\bibinfo
  {journal} {Phys. Rev. Lett.}\ }\textbf {\bibinfo {volume} {111}},\ \bibinfo
  {pages} {086601} (\bibinfo {year} {2013})}\BibitemShut {NoStop}%
\bibitem [{\citenamefont {Oreg}\ and\ \citenamefont
  {Goldhaber-Gordon}(2003)}]{31}%
  \BibitemOpen
  \bibfield  {author} {\bibinfo {author} {\bibfnamefont {Y.}~\bibnamefont
  {Oreg}}\ and\ \bibinfo {author} {\bibfnamefont {D.}~\bibnamefont
  {Goldhaber-Gordon}},\ }\href {\doibase 10.1103/PhysRevLett.90.136602}
  {\bibfield  {journal} {\bibinfo  {journal} {Phys. Rev. Lett.}\ }\textbf
  {\bibinfo {volume} {90}},\ \bibinfo {pages} {136602} (\bibinfo {year}
  {2003})}\BibitemShut {NoStop}%
\bibitem [{\citenamefont {Lebanon}\ \emph {et~al.}(2003)\citenamefont
  {Lebanon}, \citenamefont {Schiller},\ and\ \citenamefont {Anders}}]{32}%
  \BibitemOpen
  \bibfield  {author} {\bibinfo {author} {\bibfnamefont {E.}~\bibnamefont
  {Lebanon}}, \bibinfo {author} {\bibfnamefont {A.}~\bibnamefont {Schiller}}, \
  and\ \bibinfo {author} {\bibfnamefont {F.~B.}\ \bibnamefont {Anders}},\
  }\href {\doibase 10.1103/PhysRevB.68.155301} {\bibfield  {journal} {\bibinfo
  {journal} {Phys. Rev. B}\ }\textbf {\bibinfo {volume} {68}},\ \bibinfo
  {pages} {155301} (\bibinfo {year} {2003})}\BibitemShut {NoStop}%
\bibitem [{\citenamefont {Chung}\ \emph {et~al.}(2008)\citenamefont {Chung},
  \citenamefont {Zarand},\ and\ \citenamefont {W\"olfle}}]{33}%
  \BibitemOpen
  \bibfield  {author} {\bibinfo {author} {\bibfnamefont {C.-H.}\ \bibnamefont
  {Chung}}, \bibinfo {author} {\bibfnamefont {G.}~\bibnamefont {Zarand}}, \
  and\ \bibinfo {author} {\bibfnamefont {P.}~\bibnamefont {W\"olfle}},\ }\href
  {\doibase 10.1103/PhysRevB.77.035120} {\bibfield  {journal} {\bibinfo
  {journal} {Phys. Rev. B}\ }\textbf {\bibinfo {volume} {77}},\ \bibinfo
  {pages} {035120} (\bibinfo {year} {2008})}\BibitemShut {NoStop}%
\bibitem [{\citenamefont {Tanaka}\ \emph {et~al.}(2012)\citenamefont {Tanaka},
  \citenamefont {Kawakami},\ and\ \citenamefont {Oguri}}]{34}%
  \BibitemOpen
  \bibfield  {author} {\bibinfo {author} {\bibfnamefont {Y.}~\bibnamefont
  {Tanaka}}, \bibinfo {author} {\bibfnamefont {N.}~\bibnamefont {Kawakami}}, \
  and\ \bibinfo {author} {\bibfnamefont {A.}~\bibnamefont {Oguri}},\ }\href
  {\doibase 10.1103/PhysRevB.85.155314} {\bibfield  {journal} {\bibinfo
  {journal} {Phys. Rev. B}\ }\textbf {\bibinfo {volume} {85}},\ \bibinfo
  {pages} {155314} (\bibinfo {year} {2012})}\BibitemShut {NoStop}%
\bibitem [{\citenamefont {Chan}\ \emph {et~al.}(2003)\citenamefont {Chan},
  \citenamefont {Fallahi}, \citenamefont {Westervelt}, \citenamefont
  {Maranowski},\ and\ \citenamefont {Gossard}}]{35}%
  \BibitemOpen
  \bibfield  {author} {\bibinfo {author} {\bibfnamefont {I.}~\bibnamefont
  {Chan}}, \bibinfo {author} {\bibfnamefont {P.}~\bibnamefont {Fallahi}},
  \bibinfo {author} {\bibfnamefont {R.}~\bibnamefont {Westervelt}}, \bibinfo
  {author} {\bibfnamefont {K.}~\bibnamefont {Maranowski}}, \ and\ \bibinfo
  {author} {\bibfnamefont {A.}~\bibnamefont {Gossard}},\ }\href {\doibase
  https://doi.org/10.1016/S1386-9477(02)00876-7} {\bibfield  {journal}
  {\bibinfo  {journal} {Physica E: Low-dimensional Systems and Nanostructures}\
  }\textbf {\bibinfo {volume} {17}},\ \bibinfo {pages} {584} (\bibinfo {year}
  {2003})},\ \bibinfo {note} {proceedings of the International Conference on
  Superlattices, Nano-structures and Nano-devices ICSNN 2002 o-structures and
  Nano-devices ICSNN 2002}\BibitemShut {NoStop}%
\bibitem [{\citenamefont {McClure}\ \emph {et~al.}(2007)\citenamefont
  {McClure}, \citenamefont {DiCarlo}, \citenamefont {Zhang}, \citenamefont
  {Engel}, \citenamefont {Marcus}, \citenamefont {Hanson},\ and\ \citenamefont
  {Gossard}}]{36}%
  \BibitemOpen
  \bibfield  {author} {\bibinfo {author} {\bibfnamefont {D.~T.}\ \bibnamefont
  {McClure}}, \bibinfo {author} {\bibfnamefont {L.}~\bibnamefont {DiCarlo}},
  \bibinfo {author} {\bibfnamefont {Y.}~\bibnamefont {Zhang}}, \bibinfo
  {author} {\bibfnamefont {H.-A.}\ \bibnamefont {Engel}}, \bibinfo {author}
  {\bibfnamefont {C.~M.}\ \bibnamefont {Marcus}}, \bibinfo {author}
  {\bibfnamefont {M.~P.}\ \bibnamefont {Hanson}}, \ and\ \bibinfo {author}
  {\bibfnamefont {A.~C.}\ \bibnamefont {Gossard}},\ }\href {\doibase
  10.1103/PhysRevLett.98.056801} {\bibfield  {journal} {\bibinfo  {journal}
  {Phys. Rev. Lett.}\ }\textbf {\bibinfo {volume} {98}},\ \bibinfo {pages}
  {056801} (\bibinfo {year} {2007})}\BibitemShut {NoStop}%
\end{thebibliography}%
\end{document}